\documentclass[reprint,nofootinbib,superscriptaddress,amsmath,amssymb,aps,prd]{revtex4-1}
\usepackage{amsmath}
\usepackage{graphicx}
\usepackage{color}
\usepackage{dcolumn}
\usepackage{bm}
\usepackage{slashed}
\usepackage{epstopdf}
\usepackage{multirow}
\usepackage{float}
\usepackage[colorlinks,linkcolor=red,anchorcolor=green,citecolor=blue,CJKbookmarks=True]{hyperref}

\newcommand{\bea}{\begin{eqnarray*}}
\newcommand{\eea}{\end{eqnarray*}}
\newcommand{\bean}{\begin{eqnarray}}
\newcommand{\eean}{\end{eqnarray}}
\newcommand{\eqs}[1]{Eqs.(\ref{#1})}
\newcommand{\eq}[1]{Eq.(\ref{#1})}
\newcommand{\meq}[1]{(\ref{#1})}

\newcommand{\non}{\nonumber \\}

\newcommand{\hsp}{\hspace{0.1mm}}

\newcommand{\pp}{\partial}

\begin{document}
\title{Gravito-electromagnetic perturbations of MOG black holes with a cosmological constant: Quasinormal modes and Ringdown waveforms}
\author{Wentao Liu}
\affiliation{Department of Physics, Key Laboratory of Low Dimensional Quantum Structures and Quantum Control of Ministry of Education, and Synergetic Innovation Center for Quantum Effects and Applications, Hunan Normal
University, Changsha, Hunan 410081, P. R. China}

\author{Xiongjun Fang}
\email[]{fangxj@hunnu.edu.cn} \affiliation{Department of Physics, Key Laboratory of Low Dimensional Quantum Structures and Quantum Control of Ministry of Education, and Synergetic Innovation Center for Quantum Effects and Applications, Hunan Normal
University, Changsha, Hunan 410081, P. R. China}

\author{Jiliang Jing}
\affiliation{Department of Physics, Key Laboratory of Low Dimensional Quantum Structures and
Quantum Control of Ministry of Education, and Synergetic Innovation Center for Quantum Effects and Applications, Hunan Normal University, Changsha, Hunan 410081, P. R. China}

\author{Jieci Wang}
\email[]{jcwang@hunnu.edu.cn} \affiliation{Department of Physics, Key Laboratory of Low Dimensional Quantum Structures and Quantum Control of Ministry of Education, and Synergetic Innovation Center for Quantum Effects and Applications, Hunan Normal
University, Changsha, Hunan 410081, P. R. China}

\begin{abstract}
In this paper, we present a black hole solution with a cosmological constant in the Scalar-Tensor-Vector Modified Gravity (MOG) theory, where the strength of the gravitational constant is determined by $G = G_\text{N}(1+\alpha)$.
We derive the master equations for gravito-electromagnetic perturbations and numerically solve for the Quasinormal Mode (QNM) spectrum and the ringdown waveforms.
Our research results show that increasing the MOG parameter $\alpha$ leads to a decrease in both the real and imaginary parts of the QNM frequencies for electromagnetic and gravitational modes.
Similarly, increasing the cosmological constant $\Lambda$ also results in a decrease in both the real and imaginary parts of the QNM frequencies for these modes.
These trends are observed when compared to standard Schwarzschild-de Sitter (S-dS) or MOG black holes, respectively.
Meanwhile, the result indicates that in the MOG-de Sitter spacetime, the frequencies for electromagnetic and gravitational modes display isospectrality, and exhibit the same ringdown waveforms.
Our findings have implications for the ringdown phase of mergers involving massive compact objects, which is of particular relevance given the recent detections of gravitational waves by LIGO.

\end{abstract}

\maketitle

\section{Introduction}

Gravitational waves have recently opened up a new window to study fundamental questions of gravity and our universe \cite{LIGOScientific:2016sjg}.
The ringdown phase of binary black hole mergers, characterized by quasinormal modes (QNMs), provides valuable insights into the properties of the resultant black hole.
Importantly, they offer a means to constrain the charge of astrophysical black holes.
Electromagnetic fields have an impact on spacetime, leading to changes in the emission of gravitational waves compared to those from an uncharged binary system.
These deviations are accurately modeled in Einstein-Maxwell theory.
Based on this theory, some recent studies have shown that $ GW150914 $ is compatible with having charge-to-mass ratio as high as $0.3 $ \cite{Bozzola:2020mjx,Gupta:2021rod,Carullo:2021oxn}.
However, it is assumed, often implicitly, that astrophysical black holes are presumed to be electrically neutral.
This assumption is based on the principle that a black hole with mass $ M $ and electric charge $ Q $ will not gravitationally attract particles of mass $ m $ and electric charge $ e $ as long as $ eQ > Mm $.
Given that the ratio $ m/e $ is approximately $ 10^{-21} $ for electrons, it is highly unlikely for large black holes to accumulate any significant electric charge \cite{Gibbons:1975kk}.
Moreover, various mechanisms, including vacuum polarization, pair production breakdown, and neutralization from nearby material, all contribute to preventing a stellar-mass black hole from maintaining a substantial electric charge \cite{Gibbons:1975kk,Blandford:1977ds}.
Even if a significant amount of charge is obtained, the dissipation happens on a timescale that is much shorter than what can be observed through gravitational-wave studies \cite{Cardoso:2016olt}.
Therefore, we are more inclined to interpret the word ``charge'' here as dark charge, or gravitational charge \cite{Bozzola:2020mjx}.

Recently, the Scalar-Tensor-Vector Modified Gravity, proposed by Moffat \cite{Moffat:2005si}, has attached much attention. 
This theory, which introduces additional massive vector and scalar fields to the metric tensor field, assumes that the ``charge'' is a gravitational charge due to modifications to general relativity.
It has successfully explained galaxy dynamics and the cosmic microwave background power spectrum
\cite{Moffat:2013sja,Moffat:2013uaa, Moffat:2014pia,Moffat:2014bfa}.
Moreover, Moffat has obtained Schwarzschild-like and Kerr-like black hole solutions in this theory \cite{Moffat:2014aja,Lee:2017fbq,Qiao:2020fta}.
This has spurred numerous investigations into the properties of black holes within MOG theory.
For instance, research has explored the black shadow and compared the polarized images of a synchrotron emitting ring for a MOG black hole with those of M87*, finding similar spiral structures \cite{Moffat:2015kva,Guo:2018kis,Wang:2018prk,Qin2022}.
Rahvar presented the Hamiltonian formalism for the dynamics of particles and investigated the lensing on large scales and stellar size scales in \cite{Rahvar:2022yhj}. In a separate work, he also proposed a gauge-invariant theory in MOG \cite{Rouhani:2023qzy}.

Our primary objective is to extend the spacetime metric to include non-zero cosmological constants and investigate the Ringdown phase of a double MOG black hole merger.
The study of black holes with a cosmological constant is of significant importance due to their relevance to the AdS/CFT correspondence and the observation of the accelerating expansion of the universe.
Positive cosmological constants allow us to investigate astrophysical black holes that are expected to exist in our universe according to the $ \Lambda $CDM cosmological paradigm \cite{Planck:2018nkj}.
Conversely, negative cosmological constants provide an intriguing motivation to explore the QNMs of asymptotically Anti-de Sitter black holes as a means of gaining insights into specific conformal quantum field theories \cite{Maldacena:1997re,Nunez:2003eq,Son:2007vk,Hartnoll:2009sz,Herzog:2009xv}.
Under MOG, a body's mass, $M$, determines its gravitational charge, $Q$, which is linked to the vector field.
Moffat prescribes that this proportionality is set by the constant $\sqrt{\alpha G/(1+\alpha)}$, ensuring the theory adheres to the weak equivalence principle.
This hypothesis can also be applied to situations where the cosmological constant is non-zero, allowing us to obtain a black hole solution with asymptotic de Sitter and Anti de Sitter.

Furthermore, to obtain the QNM frequencies, one needs first to construct the decoupled perturbation equations in the frequency domain.
The issue of linear perturbation of black holes was pioneered by Regge and Wheeler \cite{Regge:1957td}, who presented the Regge-Wheeler formalism and considered that the metric perturbations of spherically symmetric spacetime can be decomposed into axial and polar parts.
This work was later extended by Zerilli \cite{Zerilli:1970wzz,Zerilli:1970se}. For the most general spherically symmetric metric in general relativity, the construction of decoupled equations for axial and polar gravitational perturbation have been discussed in Ref. \cite{Liu2023}.
In fact, the reliance of Regge-Wheeler formalism on spacetime symmetries allows for its application to spacetimes in general theories, such as MOG theories.
Noteworthy, the electromagnetic and gravitational perturbations of the black holes in GR possess a remarkable property that was proven by Chandrasekhar \cite{Chandrasekhar:1984siy}: the axial and polar potentials can be expressed in terms of a superpotential, implying that the polar and the axial QNMs are isospectral \cite{Berti:2009kk}.
However, there is no apparent reason for this property to hold true for MOG-dS black holes.

In this work, we are interested in the QNMs of the asymptotic de Sitter black hole with a gravitational charge.
The manuscript is organized as follows.
In Sec. \ref{Sec.2}, we briefly review the MOG theory and solve the field equations to obtain asymptotic (Anti) de Sitter solutions.
In Sec. \ref{Sec.3}, we describe how to use the Regge-Wheeler formalism for harmonic decomposition of tensor fields as well as vector fields, and derive the Schr\"{o}dinger-like equation for the gravito-electromagnetic perturbations in MOG (Anti) de Sitter black holes.
In Sec. \ref{Sec.4}, We calculate QNM frequencies and explore the effects of interaction terms on them.
In Sec. \ref{Sec.5}, the ringdown waveforms of the electromagnetic and gravitational modes are calculated using numerical methods, and the accuracy of the results is confirmed by fitting the waveforms.
Sec. \ref{Sec.6} is dedicated to summarizing our results and discussing potential extensions and future directions of research.

\section{Field equations and MOG dS black hole solution}\label{Sec.2}
In this section, we generalize the work of \cite{Moffat:2014aja} to include a cosmological constant in the STVG-MOG theory that is a covariant modified theory of gravity and whose action is composed of scalar, tensor, and vector fields \cite{Moffat:2005si}
\begin{align}
\mathcal{S}=\mathcal{S}_\text{G}+\mathcal{S}_\phi+\mathcal{S}_\text{S}+\mathcal{S}_\text{M},
\end{align}
with the parts $ \mathcal{S}_\text{G} $, $ \mathcal{S}_\phi $, and $ \mathcal{S}_\text{S} $ given by
\begin{align}
\mathcal{S}_\text{G}=&\frac{1}{16\pi}\int d^4x \sqrt{-g}\left[\frac{1}{G}\left(R-2\Lambda\right)\right],\\
\mathcal{S}_\phi=&\int d^4x \sqrt{-g}\left(-\frac{1}{4}B^{ab}B_{ab}+\frac{1}{2}\mu^2\phi^a\phi_a\right),\\
\mathcal{S}_\text{S}=&\int d^4x\sqrt{-g}\left[\frac{1}{G^3}\left(\frac{1}{2}g^{ab}\nabla_a G\nabla_b G -V(G)\right)
\right.\non&\left.
+\frac{1}{\mu^2 G}\left(\frac{1}{2}g^{ab}\nabla_a\mu\nabla_b\mu-V(\mu)\right)\right].
\end{align}
Here, $\mathcal{S}_\text{G}$ is the Einstein-Hilbert action, $\Lambda$ and $R$ denote the cosmological constant and the Ricci scalar, respectively. $\mathcal{S}_\phi$ is the action of $\phi^a$ that is a Proca-type massive vector field with mass $\mu$, while $B_{ab}$ is its field strength, defined as $ B_{ab}=\pp_a\phi_b-\pp_b\phi_a $, which satisfies the following equations,
\begin{align}
\pp_cB_{ab}+\pp_aB_{bc}+\pp_bB_{ca}=0.
\end{align}
$S_\text{S}$ contains self-interaction potentials $ V(G) $ and $V(\mu) $, which correspond to the scalar fields $ G(x) $ and $ \mu(x) $, respectively. 
The action $ \mathcal{S}_\text{M} $ represents matter, with its current density $ J^a $ defined as:
\begin{align}
\frac{1}{\sqrt{-g}}\frac{\delta \mathcal{S}_\mathrm{M}}{\delta \phi_a}=-J^a,
\end{align}
illustrating the interaction between matter and the vector field \cite{Moffat:2005si}.
This suggests every particle carries an extra charge proportional to its inertial mass and $J^a=\kappa \rho u^a $, as detailed in Ref. \cite{Rahvar:2022yhj}.
The parameter $ G $ is a scalar field that corresponds to a spin-$ 0 $ massless graviton, which is related to Newton's gravitational constant $ G=G_\text{N}(1+\alpha) $, where $\alpha$ is a dimensionless parameter and the best fit to the spiral galaxies results in $ \alpha=8.89\pm 0.34 $ \cite{Moffat:2013sja}.
The MOG theory would back to GR when $\alpha=0$. As a result, we can regard $ \alpha $ as a deviation parameter of the MOG from GR.

Given that $G$ is considered a constant independent of the spacetime coordinates, using a vacuum solution will simplify the action to 
\begin{equation}
\begin{aligned}\label{action}
\mathcal{S}=&\int d^4x\sqrt{-g}\Big[\frac{1}{16\pi G}\left(R-2\Lambda\right)\\
&-\frac{1}{4}B^{ab}B_{ab}+\frac{1}{2}\mu^2\phi^a\phi_a \Big]+\mathcal{S}_\mathrm{M}.
\end{aligned}
\end{equation}
The field equation of motion derived from varying the action in Eq. \meq{action} with respect to the metric is given by
\begin{align}\label{gravityEq}
\mathcal{G}_{ab}=G_{ab}+\Lambda g_{ab}+\frac{8\pi G}{c^4} T^\phi_{ab}=0,
\end{align}
with the energy momentum tensor given by
\begin{align}
\begin{aligned}
T^{\phi}_{ab}=&-\frac{1}{4\pi}\Big(B_a\hsp^cB_{bc}-\frac{1}{4}g_{ab}B^{cd}B_{cd}\Big)\\
&+\frac{\mu^2}{4\pi}\Big( \phi_a\phi_b-\frac{1}{2}g_{ab}\phi^c\phi_c \Big).
\end{aligned}
\end{align}
The first and second terms of the energy-momentum tensor are of the order $\sim(\pp \phi)^2$ and $\sim\mu^2 \phi^2$, respectively.
At the same time, we vary the action in equation \meq{action} with respect to $\phi^a $, obtaining:
\begin{align}\label{Pia}
\Pi^a=\nabla_bB^{ab}-\mu^2\phi^a+4\pi J^a=0.
\end{align}
For a distribution of matter, the non-zero source of current is given by $ J^0=\kappa \rho $  \cite{Rahvar:2022yhj}.

In this work, we examine the static gravitational field where $\phi^a $ has a zero time derivative and adopt the following static spherically symmetric metric
\begin{align}
ds^2=-F(r)dt^2+F(r)^{-1}r^2+r^2d\Omega^2,
\end{align}
where $ d\Omega^2=d\theta^2+\sin^2\theta d\varphi^2 $. 
Considering the vector field $\phi^a$ has the form $\phi^a=(\phi^0(r),0,0,0)$, we can get all non-zero components equations as follows:
\begin{align}\label{eqM0}
\Pi^0=&\nabla^2\Phi+\frac{\mu^2}{F}\Phi+4\pi J^0,\\ \label{eqE0}
\mathcal{G}_{00}=&G\left(\mu^2\Phi^2-F{\Phi'}^2\right)+\frac{F(1-F-rF'-r^2\Lambda)}{r^2},\\ \label{eqE1} 
\mathcal{G}_{11}=&\frac{G}{F^2}\left(\mu^2\Phi^2+F{\Phi'}^2\right)-\frac{(1-F-rF'-r^2\Lambda)}{r^2F},\\ \label{eqE2}
\mathcal{G}_{22}=&\frac{Gr^2}{F}\left(\mu^2\Phi^2-F{\Phi'}^2\right)+\frac{1}{2}r^2F''+r^2\Lambda+rF',\\ \label{eqE3}
\mathcal{G}_{33}=&\sin^2\theta \mathcal{G}_{22},
\end{align}
where $ \Phi(r)=-F(r) \phi^0(r) $.

By combining equations \meq{eqE0} and \meq{eqE1}, it is found that the system of differential equations is subject to the following constraints,
\begin{align}
G\mu^2\Phi^2=0.
\end{align}
This implies that we must address the problem under certain approximations. 
In the common weak field approximation, higher-order terms such as $ g^2 $, $ g\phi $ and $ \phi^2 $ are typically neglected \cite{Moffat:2013sja,Rahvar:2022yhj}. 
Concurrently, according to Refs. \cite{Moffat:2013sja,Moffat:2013uaa}, the particle mass of the $ \phi^a $ field in the present universe can be fitted as $ m_\phi\sim 10^{-28}eV $, making it negligible for a black hole solution.
In this paper, we choose to neglect the $ \phi^a $ field particle mass. 
Consequently, the solution of \eq{eqM0} can be given by
\begin{align}
\Phi(x)=\int \frac{\kappa \rho(\mathbf{x}')}{|\mathbf{x}-\mathbf{x}'|}d^3\mathbf{x}'.
\end{align}
Using the Dirac-delta function $\rho(\mathbf{x}')=M\delta^3(\mathbf{x}')$, the potential reduces to
\begin{align}
\Phi=\frac{\kappa  M}{r}.
\end{align}
In this theory, it's crucial to note that the fifth force charge, $Q_5  $, is posited to be proportional to the inertial mass of a particle, expressed as $ Q_5=\kappa M $ \cite{Rahvar:2022yhj,Moffat:2013sja}.
Subsequently, the corresponding metric function can be obtained by solving Eq. \meq{eqE0} or Eq. \meq{eqE2}, as
\begin{align}
F=&1-\frac{2GM}{r}+\frac{GQ_5^2}{r^2}-\frac{\Lambda }{3}r^2.
\end{align}
Using the convention of $ \kappa^2 =\alpha G_\text{N} $ \cite{Rahvar:2022yhj}, we can rewrite the solution as
\begin{align}\label{phi0}
\phi^0=&-\sqrt{\alpha G_\text{N}}\frac{M}{rF}.
\end{align}

For simplicity, we set $ G_\text{N}=1 $ in next discussion, hence the metric can be written as
\begin{align}
\begin{aligned}\label{ds2MOG}
ds^2=&-\left(1-\frac{2M_\text{D}}{r}+\frac{\beta M^2_\text{D}}{r^2}-\frac{\Lambda}{3}r^2\right)dt^2
\\&
+\left(1-\frac{2M_\text{D}}{r}+\frac{\beta M^2_\text{D}}{r^2}-\frac{\Lambda}{3}r^2\right)^{-1}dr^2
+d\Omega^2,
\end{aligned}
\end{align}
where $ \beta=\alpha/(1+\alpha) $, and $ M_\text{D} $ is the ADM mass \cite{Sheoran:2017dwb}, which is related to the Newtonian mass $M$ as $M_\text{D}=(1+\alpha)M$.
Note that this solution appears to have the same form as the RN-(A)dS solution. However, the charge $Q_5=\sqrt{\beta G}M$ is of gravitational origin, rather than electric charge.
If $ \alpha \rightarrow0 $, it recovers the usual S-dS metric.
When $\Lambda\rightarrow 0$ , the above solution becomes Eq. (14) of Ref. \cite{Moffat:2014aja}. When $\Lambda>0$ or $\Lambda<0$, the solution describes the MOG-de Sitter (MOG-dS) black hole or the MOG Anti-de Sitter (MOG-AdS) black hole, respectively.

The horizon surface equation of the spacetime is
\begin{align}
F=\frac{\Lambda}{3}\left(1-\frac{r_m}{r}\right)\left(1-\frac{r_h}{r}\right)\left(r_c-r\right)\left(r+r_b\right)=0,
\end{align}
where $r_m$, $r_h$ and $r_c$ represent the inner event horizon, the outer event horizon and the cosmological horizon, respectively.
And $r_b$ can be determined by the relation $ r_b+r_m+r_h+r_c=0. $

\section{Gravito-electromagnetic perturbations}\label{Sec.3}
\subsection{ Harmonic decomposition}
Using $h_{ab}$ and $\delta\phi_a$ to represent the linear perturbation of the background metric $ g^{(0)}_{ab} $ and the vector field $ \phi_a^{(0)} $, respectively, then the perturbed spacetime and its field can be written as
\begin{align}
g_{ab}=g^{(0)}_{ab}+h_{ab},&& \phi_a=\phi^{(0)}_a+\delta \phi_a.
\end{align}
We decompose the metric perturbations $ h_{ab} $ in the Regge-Wheeler gauge \cite{Regge:1957td,Zerilli:1970wzz,Zerilli:1970se,Liu2023,Zhao:2023jiz}:
\begin{align}\label{deltametric}
h_{ab}=\left(\begin{array}{cccc}
H_0Y^{lm} &H_1 Y^{lm}  &h_0 S^{lm}_{\theta}  &h_0S^{lm}_{\varphi}  \\
Sym& H_2Y^{lm} & h_1S^{lm}_{\theta} &h_1S^{lm}_{\varphi}  \\
Sym &Sym  &r^2 KY^{lm}  & 0 \\
Sym &Sym  & Sym &r^2\sin^2\theta K  Y^{lm}
\end{array}\right),
\end{align}
where $ Y^{lm}=Y^{lm}\left(\theta,\varphi\right) $ is the ordinary scalar spherical harmonics.
$S^{lm}_{\theta}:=-\csc\theta \frac{\pp}{\pp\varphi }Y^{lm}$ and $S^{lm}_{\varphi}:= \sin\theta \frac{\pp}{\pp \theta}Y^{lm}$ are the axial vector harmonics, since under parity transformation $ \left(\theta,\varphi\right)\rightarrow\left(\pi-\theta ,\pi+\varphi\right)$ these modes pick a factor $ \left(-1\right)^{l+1} $.
Meanwhile, we expand the perturbation of the vector field $\delta\phi_a$ as follows \cite{Rosa:2011my,Zhang:2023wwk}:
\begin{align}\label{deltavector}
\delta \phi_{a}=
\left[
\begin{array}{c}
0 \\ 0  \\ u_{(4)} S^{lm}_b/\lambda
\end{array}
\right]
+\left[
\begin{array}{c}
u_{(1)}Y^{lm} /r\\u_{(2)}Y^{lm}/\left(rF\right)  \\ u_{(3)} Y^{lm}_b/\lambda
\end{array}
\right],
\end{align}
where $\lambda=l(l+1)$, $b=\left(\theta,\varphi\right)$, $Y^{lm}_{b}:=\frac{\pp}{\pp b}Y^{lm}$ are the polar vector harmonics for the pick a factor $ \left(-1\right)^{l} $.
Note that, the spherical harmonic function $ Y^{lm} $ is part of the polar sector and all perturbation functions $h_{0,1}, H_{0,1}, K $ and $u_{(1,2,3,4)}$ are functions of $(t,r)$.
Inserting the harmonic expansion of the metric perturbation \meq{deltametric} and the vector perturbation \meq{deltavector} into the linearized field equations \meq{gravityEq} and \meq{Pia}, one can obtain all components of the equations.
In fact, there is no reason to consider that the axial and polar parts will not be automatically separated \cite{Pani:2013ija,Pani:2013wsa,Nomura:2020tpc,Meng:2022oxg,Guo:2022rms}.
However, by separating the angular dependence \cite{Thorne:1980ru}, a system of fourteen coupled pure radial equations can be obtained. These equations consist of ten gravitational sector and four Maxwell sector, which are naturally separated into axial parity and polar parity.

We perform a Fourier decomposition by assuming that all perturbations have a time dependence $\sim e^{-i\omega t}$.
Then, all ordinary differential equations (ODEs) are listed in Appendix \ref{Appendixeq}. Note that in electromagnetic perturbations, we make the assumption that the perturbation of the current has a linear relationship with the perturbation of the additional vector field, as given by $ \delta J^a=-\frac{\xi}{4\pi r}\frac{\pp }{\pp r}\left(F \delta \phi^a\right)$. 
Here, the parameter $ \xi $ can either be $ 1 $ or $ 0 $, which correspond to the inclusion or exclusion of the interaction term perturbation, respectively.
In the next subsection, we aim to obtain the decoupled equations for the electromagnetic and gravitational fields, respectively \cite{Zerilli:1974ai,Moncrief:1974am,Moncrief:1975sb}.

\subsection{Derivation of the perturbation equations}\label{32}
\subsubsection{Axial sector}
In Appendix \ref{Appendixeq}, the Eqs. \meq{EQG4}, \meq{EQG7}, \meq{EQG8}, and \meq{EQE4} form a set of coupled systems for axial sector.
However, it is worth noting that Einstein's equations implies Maxwell's equations, and thus only three out of the four equations listed above are independent.
These three independent equations can be solved for the functions $h_{0,1}$, and $u_{(4)}$.
We can define the Regge-Wheeler functions as follows:
\begin{align}
\psi_g= \frac{F}{r} h_1,&& \psi_e=-r^{\xi/2} u_{(4)},
\end{align}
and by utilizing Eq. \meq{EQG8}, the perturbation function $h_0$ can be eliminated.
Subsequently, through some straightforward algebraic manipulation, we arrive at a system of coupled second-order equations for $ \psi_e $ and $\psi_g$ , as
\begin{equation}\label{odd}
\begin{aligned}
&\frac{d^2}{dr^2_*}\psi_g+\left(\omega^2-\mathcal{V}_1\right)\psi_g=\frac{4i\omega M\sqrt{\alpha}(1+\alpha)F}{r^{3+\xi/2}\lambda} \psi_e,\\
&\frac{d^2}{dr^2_*}\psi_e+\left(\omega^2-\mathcal{V}_2\right)\psi_e=-\frac{iM\sqrt{\alpha}\lambda}{r^{4-\xi/2}\omega}\mathcal{D}_2 \psi_g,
\end{aligned}
\end{equation}
where $ r_* $ is the tortoise coordinate defined by $ dr/dr_*=F $, and
\begin{equation}
\begin{aligned}
\mathcal{V}_1=&\frac{2F^2}{r^2}-\frac{F}{r^2}\left(2-\lambda+rF'\right),\\
\mathcal{V}_2=&\frac{F}{4r^2}\left[4\lambda+\xi(6+\xi)F-2\xi rF'\right]\\
&-\frac{4F}{r^4}(\xi-1)M^2\alpha(1+\alpha),
\\ \mathcal{D}_2=&\xi r^3\omega^2+(\xi-1)(2-\lambda)rF+\xi r F^2(1+ \frac{d}{dr}).
\end{aligned}
\end{equation}
Now this system of coupled equations can be solved by using numerical methods. For the case $ \xi=0 $, it can further decouple the system. 
This involves defining a linear combination of two new functions, $ Z^{(-)}_i(i=1,2) $, such that
\begin{equation}
\begin{aligned}
Z^{(-)}_1=\frac{(\sigma+3+3\alpha)}{2\sigma}\psi_e+\frac{\sqrt{\alpha}\lambda(\lambda-2)}{2i\omega\sigma}\psi_g,\\
Z^{(-)}_2=\frac{(\sigma-3-3\alpha)}{2\sigma}\psi_e-\frac{\sqrt{\alpha}\lambda(\lambda-2)}{2i\omega\sigma}\psi_g,
\end{aligned}
\end{equation}
with
\begin{align}
\sigma=\sqrt{(1+\alpha)(9+\alpha+4\alpha \lambda)}.
\end{align}
By substituting the aforementioned linear combinations into the equations for $\psi_e $ and $\psi_g$, and solving for $Z^{(-)}_1$ and $Z^{(-)}_2$, we obtain the following expressions:
\begin{align}
\frac{d^2}{dr^2_*}Z^{(-)}_i+\left[\omega-V^{(-)}_{i}\right]Z^{(-)}_i=0,
\end{align}
where
\begin{equation}
\begin{aligned}
V^{(-)}_i=\frac{F}{r^4}\left[r^2\lambda+M(1+\alpha)(4M\alpha-3r)-(-1)^{i}rM\sigma\right].
\end{aligned}
\end{equation}
In the limit $\alpha=0$, the effective potentials $V^{(-)}_1=\mathcal{V}_1$ and $V^{(-)}_2=\mathcal{V}_2$ reduce to the corresponding potentials for axial electromagnetic and gravitational perturbations of S-(A)dS black holes \cite{Cardoso:2001bb,Zhidenko:2003wq}, respectively.
Therefore, for the sake of convenience, we will refer to these two modes as the ``electromagnetic" mode and the ``gravitational" mode even in the general case.
It should be emphasized that when $ \alpha\neq0 $, oscillations involving either of these modes excite both electromagnetic and gravitational perturbations.

\subsubsection{Polar sector}


In Appendix \ref{Appendixeq}, the remaining ten equations constitute a set of coupled systems for the polar sector.
The Bianchi identities reveal that not all equations are independent.
In fact, only seven equations are truly independent and they can be solved for the seven polar functions: $ H_{0,1,2}, K$  and $ u_{(1,2,3)} $. Note that in this subsection, it becomes impossible to obtain the master perturbation equation when we choose $ \xi=1 $. 
Therefore, we only consider the case $ \xi=0 $.
The key point in constructing the master equation for the gravito-electromagnetic perturbation is to use the perturbation of the field strength $ B_{ab} $ as the dynamical variable, instead of the vector potential $\delta \phi_{a} $.
Following the approach in \cite{Zerilli:1974ai}, we define
\begin{align}\label{Fab}
\delta B_{ab}:=f_{ab}=\pp_a\delta \phi_{b}-\pp_b\delta \phi_{a}.
\end{align}
In the polar sector, we can fix the gauge by requiring $ u_{(3)}=0 $ based on the Lorentz condition.
the remaining components are related to $ f_{ab} $ in the following way:
\begin{align}
u_{(1)}&=r\tilde{f}_{02},\\
u_{(2)}&=r F(r)\tilde{f}_{12},\\
u_{(1)}'&=r\tilde{f}_{01}+\tilde{f}_{02}-i\omega r \tilde{f}_{02}
\end{align}
where $-\tilde{f}_{ab} $ denotes the angle-independent part of $ f_{ab} $,
and the homogeneous Maxwell equation
\begin{align}\label{homomaxwell}
\tilde{f}_{01}=\tilde{f}'_{02}+i\omega \tilde{f}_{12}
\end{align}
is automatically satisfied.

First, we solve Eq. \meq{EQG9} for $ H_2 $ and substitute the result into the remaining equations.
Then, by solving for $ \tilde{f}{01} $ and $ \tilde{f}{02} $ using Eqs. \meq{EQE2}-\meq{EQE3} and incorporating Eq. \meq{homomaxwell}, we can obtain a second-order differential equation in the following form:
\begin{align}\label{EQEM1}
\frac{d^2}{dr_*^2}f_{EM}+\left(\omega^2-\frac{\lambda F}{r^2}\right)f_{EM}=\frac{i\omega\sqrt{\alpha}MF}{r^2}K,
\end{align}
where
\begin{align}
f_{EM}=F \tilde{f}_{12}.
\end{align}

To derive the equations for the gravitational sector, we solve Eqs. \meq{EQG2}, \meq{EQG3}, and \meq{EQG6} for $ H_0' $, $ H_1' $, and $ K' $.
After that, substituting this solution into Eq. \meq{EQG5}, we solve the function $ H_0 $ and eliminate it in the remaining equations.
And define
\begin{align}
R=\frac{1}{\omega}H_1 ,
\end{align}
as a result of this procedure we obtain a system of coupled equations
\begin{align}
K'=&\left(\alpha_0+\alpha_2\omega^2\right)K+\left(\beta_0+\beta_2\omega^2\right)R+S_1,\\
R'=&\left(\gamma_0+\gamma_2\omega^2\right)K+\left(\delta_0+\delta_2\omega^2\right)R+S_2,
\end{align}
where $ \alpha_{0,2},\beta_{0,2},\gamma_{0,2},\delta_{0,2} $ are coefficients that do not depend on $ \omega $. The source terms $ S_1 $ and $ S_2 $ consist of the perturbation functions $ f_{EM} $ and $ f_{EM}' $.

Following Zerilli's approach \cite{Zerilli:1970se,Zerilli:1974ai}, we assume the transformation as
\begin{equation}\label{ZerilliT}
\begin{aligned}
&K=f(r)\hat{K}+g(r)\hat{R},&&R=h(r)\hat{K}+k(r)\hat{R},
\end{aligned}
\end{equation}
and then obtain a Schr\"{o}dinger-type equation,
\begin{align}\label{Zerilli}
\frac{d}{d\hat{r}}\hat{K}=\hat{R}+\hat{S}_1,&& \frac{d}{d\hat{r}}\hat{R}=\left[\mathcal{U}_2-\omega^2\right]\hat{K}+\hat{S}_2.
\end{align}
here $\hat{r}$ is determined by $dr/d\hat{r}=n(r)$, and $ f,g,h $ and $ k $ are given by
\begin{equation}
\begin{aligned}
&f(r)=\frac{\lambda}{2r}-\frac{4c_1+r\lambda F'}{4\varpi},&&g(r)=1,\\
&h(r)=i-\frac{4rc_2+r^2\lambda F'}{4iF \varpi}, &&k(r)=\frac{r}{i F},\\
&n(r)=F,
\end{aligned}
\end{equation}
where $ \varpi $, $ c_1 $ and $ c_2 $ are functions determined by the background and can be find in Appendix \ref{AppendixCCC}. Thus the variable $ \hat{r} $ is just the variable $ r_* $.

From \eq{Zerilli}, a single variable second-order equation for $ \hat{K} $ can be written as
\begin{align}\label{couplingG}
\frac{d^2}{d r_*^2}\hat{K}+\left(\omega^2-\mathcal{U}_2\right)\hat{K}=\frac{\sqrt{\alpha}(1+\alpha)}{i\omega r^2}\mathcal{W}f_{EM}.
\end{align}
And the effective potential for gravitational perturbations is
\begin{align}
\mathcal{U}_2=&\frac{F}{12r^6\varpi^2}\left\{48M^5\alpha^2(1+\alpha)^3(9r-2M\alpha)\right.\non
&-24M^4r^2\alpha (1+\alpha)^2[27+\alpha (33-3\lambda+4r^2\Lambda)]\non
&+24M^3r^3(1+\alpha)^2[9+\alpha(31-8\lambda+6r^2\Lambda)]   \non
&+36M^2r^4(1+\alpha)(3+5\alpha)(\lambda-2)\non
&-8M^2r^6(1+\alpha)(9+7\alpha+\lambda\alpha)\Lambda\non
&\left.+3r^5(\lambda-2)^2[6M(1+\alpha)+r\lambda] \right\},
\end{align}
and the parameter
\begin{equation}
\begin{aligned}
\mathcal{W}=&\frac{2MF}{3r^2\varpi^2}\left\{4Mr^2(1+\alpha)[9+r\Lambda(3r-4M\alpha)]\right.\\
&\left.-12M^2(1+\alpha)^2(3r-M\alpha)+3r^3(\lambda^2-4)\right\}.
\end{aligned}
\end{equation}

Now we focus on the electromagnetic sector once again. Using Eqs. \meq{ZerilliT} and \meq{Zerilli} to solve $K$ and substitute it into the right-hand side of Eq. \meq{EQEM1}, we obtain:
\begin{align}\label{couplingE}
\frac{d^2}{dr^2_*}f_{EM}+\left(\omega^2-\mathcal{U}_1\right)f_{EM}=\hat{D}\hat{K},
\end{align}
where
\begin{align}
\hat{D}=\frac{i\omega \sqrt{\alpha}MF}{r^2}[f(r)+\frac{d}{dr_*}],
\end{align}
and the effective potential for electromagnetic perturbations is
\begin{align}
\mathcal{U}_1=\frac{\lambda F}{r^2}+\frac{4M^2\alpha(1+\alpha)F^2}{r^3\varpi}.
\end{align}
Similar to the axial parity case, we end up with a system of coupled second-order equations for $ f_{EM} $ and $ \hat{K} $.

Finally, it is possible to decouple these equations by introducing the functions $ Z^{(+)}_i $ such that
\begin{equation}
\begin{aligned}
f_{EM}&=\mathcal{B}_{11}Z^{(+)}_2+\mathcal{B}_{12}Z^{(+)}_1,\\
\hat{K}&=\mathcal{B}_{21}Z^{(+)}_2+\mathcal{B}_{22}Z^{(+)}_1,
\end{aligned}
\end{equation}
where $ \mathcal{B}_{ij} $ is either a constant or a function of $ r $.
It is not difficult to verify that we have the following choices
\begin{equation}
\begin{aligned}
\frac{\mathcal{B}_{11}}{C_1}+\sigma=\frac{\mathcal{B}_{12}}{C_2}-\sigma=&\frac{1}{r}(1+\alpha)\left(3r-4M\alpha\right),\\
\frac{\mathcal{B}_{21}}{C_1}=\frac{\mathcal{B}_{22}}{C_2}=&-\frac{8i\sqrt{\alpha}(1+\alpha)}{\omega},
\end{aligned}
\end{equation}
where $ C_1 $ and $ C_2 $ are integration constants that can be set to unity without loss of generality.
After solving for $ Z^{(+)}_1 $ and $ Z^{(+)}_2 $ by associating Eqs. \meq{couplingG} and \meq{couplingE}, the final equations take the following form
\begin{align}
\frac{d^2}{dr_*^2}Z^{(+)}_i+\left[\omega^2-V^{(+)}_i\right]Z^{(+)}_i=0,
\end{align}
and the potentials appearing in the polar sector of the perturbation equations are given by
\begin{equation}
\begin{aligned}
V^{(+)}_i=&\frac{\mathcal{U}_1+\mathcal{U}_2}{2}-\frac{(-1)^i}{2r^4\sigma}\left\{\alpha(1+\alpha) r\varpi \mathcal{W}\right.\\
&-r^3(1+\alpha)(3r-4M\alpha)\left(\mathcal{U}_2-\mathcal{U}_1\right)\\
&\left.+4rM\alpha(1+\alpha) F\left[2F-rF'+2rf(r)\right]\right\}.
\end{aligned}
\end{equation}
Furthermore, when $ \alpha\rightarrow0 $, the potential $ V^{(+)}_1=\mathcal{U}_1 $ and $ V^{(+)}_2=\mathcal{U}_2 $ reduce to the potential for polar electromagnetic and gravitational perturbations of S-(A)dS, respectively, i.e. the Eqs. (6) and (8)  in Ref. \cite{Zhidenko:2003wq}.
In the next section, using the matrix method and the WKB approach, our results show how the parameter $\alpha$ and the cosmological constant $\Lambda$ affect the QNM frequencies.

\section{ Quasinormal Modes}\label{Sec.4}
\subsection{Case for $ \xi=0 $}\label{QNMs0}

In this subsection, the focus is on the QNM spectrum of MOG-dS black holes and their isospectrality. 
Using the matrix method proposed by Lin et al. \cite{Lin:2016sch,Lin:2017oag,Lin:2019mmf,Lei:2021kqv,Liu:2022dcn}, the QNMs can be calculated.
The gravito-electromagnetic perturbation equations can be uniformly written as
\begin{align}\label{mastereq}
\frac{d^2}{d r_*^2}\psi+\left[\omega^2-V(r)\right]\psi=0,
\end{align}
The equations satisfy two boundary conditions, owing to the existence of the two horizons.
The generic wave function $\psi$ has the asymptotic behaviors as
\begin{equation}\label{solveeq}
\psi\sim
\begin{cases}
e^{-i\omega r_*} &\text{for}~~r\rightarrow~r_h,\\
e^{i\omega r_*} &\text{for}~~r\rightarrow~r_c .
\end{cases}
\end{equation}
To obtain the QNMs with radial interval $r_h\leq r\leq r_c$, the tortoise coordinate $r_*$ can be rewritten as
\begin{equation}
\begin{aligned}
r_*=&\eta_m  \ln \left(r-r_m \right)+\eta_h\ln\left(r-r_h\right)
\\&
+\eta_c \ln \left(r_c-r\right)+\eta_b\ln\left(r+r_m+r_h+r_c\right),
\end{aligned}
\end{equation}
where
\begin{align}
\eta_i=&\frac{3r_i}{\Lambda}\left(r_{j_1}-r_i\right)^{-1}\left(r_{j_2}-r_i\right)^{-1}\left(r_{j_3}-r_i\right)^{-1},
\end{align}
and  $ i\neq j_1\neq j_2 \neq j_3 $. To convert the radial interval into $ [0,1] $, we introduce a coordinate transformation
\begin{align}
x=\frac{r-r_h}{r_c-r_h}.
\end{align}

Together with the asymptotic solutions \meq{solveeq}, we consider that $\psi$ satisfied the relation as
\begin{align}\label{psi}
\psi=(1-x)^{i \omega \eta_c}x^{- i\omega \eta_h} \mathcal{R}(x).
\end{align}
This implies that $ \mathcal{R}(0)=\mathcal{R}_0 $ and $ \mathcal{R}(1) = \mathcal{R}_1 $, where $ \mathcal{R}_0 $ and $ \mathcal{R}_1 $ are indeterminate constants.
Again, considering the boundary condition, we further introduce
\begin{align}
\chi(x)=x(1-x)\mathcal{R}(x).
\end{align}
This boundary condition ensures that $ \chi(0)=\chi(1)=0 $, and the resulting matrix equation is homogenous. Then, the perturbation equations for all effective potentials can be rewritten in the following form
\begin{align}\label{qicieq}
\mathcal{C}_2(x,\omega)\chi''(x)+\mathcal{C}_1(x,\omega)\chi'(x)+\mathcal{C}_0(x,\omega)\chi(x)=0.
\end{align}
Numerically, we set $ M = 1 $, where $\mathcal{C}_j(j=0,1,2)$ is determined by the effective potentials, the modes, and the black hole horizons $ r_m $, $ r_h $ and $ r_c $ which are determined by the black hole parameters $ \alpha $ and $ \Lambda $.

The matrix method is a non-grid-based interpolation approach.
We need to discretize Eq. \meq{qicieq} and introduce equally spaced grid points into the internal $[0,1]$.
The corresponding differential matrices can be constructed by expanding the function $ \chi(x) $ around each grid point using the Taylor series.
Thus, the differential Eq. \meq{qicieq} is therefore rewritten as an algebraic equation, as
\begin{align}\label{algeq}
\left(\mathcal{M}_0+\omega\mathcal{M}_1\right)\chi(x)=0,	
\end{align}
where $ \mathcal{M}_0 $ and $ \mathcal{M}_1 $ are matrices consisting of the functions $\mathcal{C}_j$ and the corresponding differential matrices, which allows us to easily obtain the QNM frequencies.

For comparison, here we also use the sixth-order WKB approach to calculate the QNM frequencies, which is analogous to the problem of waves scattering near the peak of the potential barrier $ V $ in quantum mechanics \cite{Schutz:1985km}.  For general potential $ V $, the formula to get $\omega$ in six-order WKB approach is given by
\begin{align}
i\frac{\omega^2-V_0}{\sqrt{-2V_0''}}-\Lambda_2-\Lambda_3-\Lambda_4-\Lambda_5-\Lambda_6=n+\frac{1}{2},
\end{align}
where $V_0=V|_{r_*=r_\text{max}}$, $V_0''=\frac{d^2}{dr_*^2}V|_{r_*=r_\text{max}}$, and the specific form of the terms  $ \Lambda_{i} (i=2,3,4,5,6) $ can be found in \cite{Iyer:1986np,Konoplya:2003ii}.
Generally speaking, the accuracy of the matrix method depends on the number of grid points, while the accuracy of the WKB approach depends on the order of the correction terms $ \Lambda_{i} $. We provide numerical results by these two ways to provide validation for each other.

In the content that follows, we show the results of numerical results of QNM frequencies by using the matrix method or the WKB approach. In the matrix method, comparing the results of setting the grid points at $ 20 $ or $ 40 $, we observe that the percentage error between the cases is less than $10^{-5}$.
As a result, we consider that setting the number of grid points $N=20$ would provide sufficient accuracy.
Then, we present data only for some of the lower modes as the higher modes produce larger values for the imaginary part of the frequency and these modes have a faster decay rate compared to the lower modes.
It is expected that low-frequency modes will be the most significant in astrophysical applications, which are most relevant to gravitational wave detection \cite{Berti:2005ys}.
None of our numerical searches (for $0< r_h<r_c$, and $ l=1,2,3 $) returned exponentially growing modes.
The calculation data as a reference can be found in the Appendix \ref{AppendixQNM}.

\subsubsection{The electromagnetic modes}\label{EMmode}
The electromagnetic modes exist for $ l\geq 1 $.
For specificity, our focus will be on the axial electromagnetic modes with azimuthal indices of $ l = 1 $ and $ l = 2 $.
When $\Lambda=0$, using the asymptotic iteration method (AIM) \cite{Ciftci:2005xn}, Manfredi calculated the QNM frequencies for electromagnetic perturbations of the MOG black hole \cite{Manfredi:2017xcv}.
It is shown in Tab. \ref{tab:1} that the results from the WKB approach are consistent with those from the AIM.
\begin{table}[h]
\renewcommand{\arraystretch}{1.2}
\centering
\caption{Contrasting the axial complex electromagnetic frequencies for values of $ M = 1 $ and $ \alpha = 1 $ in MOG-black holes.}
\label{tab:1}
\setlength\tabcolsep{2mm}{
\begin{tabular}{cccccc}
\hline\hline
\multirow{2}{*}{~\bf }	&
\multirow{1}{*}{~\bf $ n $~}  &
\multicolumn{1}{c}{WKB approach} & \multicolumn{1}{c}{Manfredi}
\\
\hline
\multirow{1}{*}{$ l=1 $}
  &0  &0.1504-0.04890i 	&0.1448-0.04805i
\\
\\
\multirow{2}{*}{$ l=2 $}
  &0  &0.2693-0.04941i	&0.2651-0.04917i
\\&1  &0.2609-0.1504i 	&0.2565-0.1498i	
\\
\hline\hline
\end{tabular}}
\end{table}

Now we turn to the asymptotically de Sitter spacetime withp non-vanishing $ \Lambda $.
When $ \alpha=0 $, the QNMs have been calculated by Zhidenko \cite{Zhidenko:2003wq} and using the matrix method by Lin et al. \cite{Lin:2016sch}.
Here we consider the parameter $\alpha$ in the range $\alpha \in (0,1]$, since it requires the simultaneous existence of both the black holes outer event horizon and the cosmological horizon in de Sitter spacetime.
\begin{figure}[h]
\centering
\includegraphics[width=0.48\linewidth]{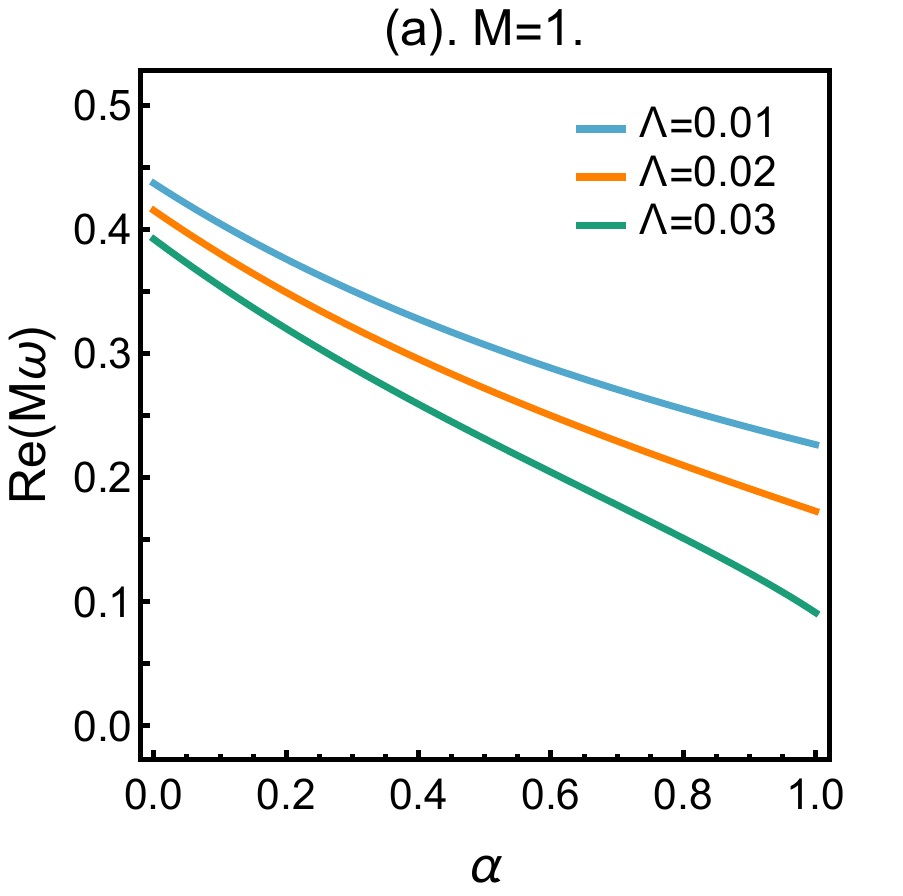}
\includegraphics[width=0.48\linewidth]{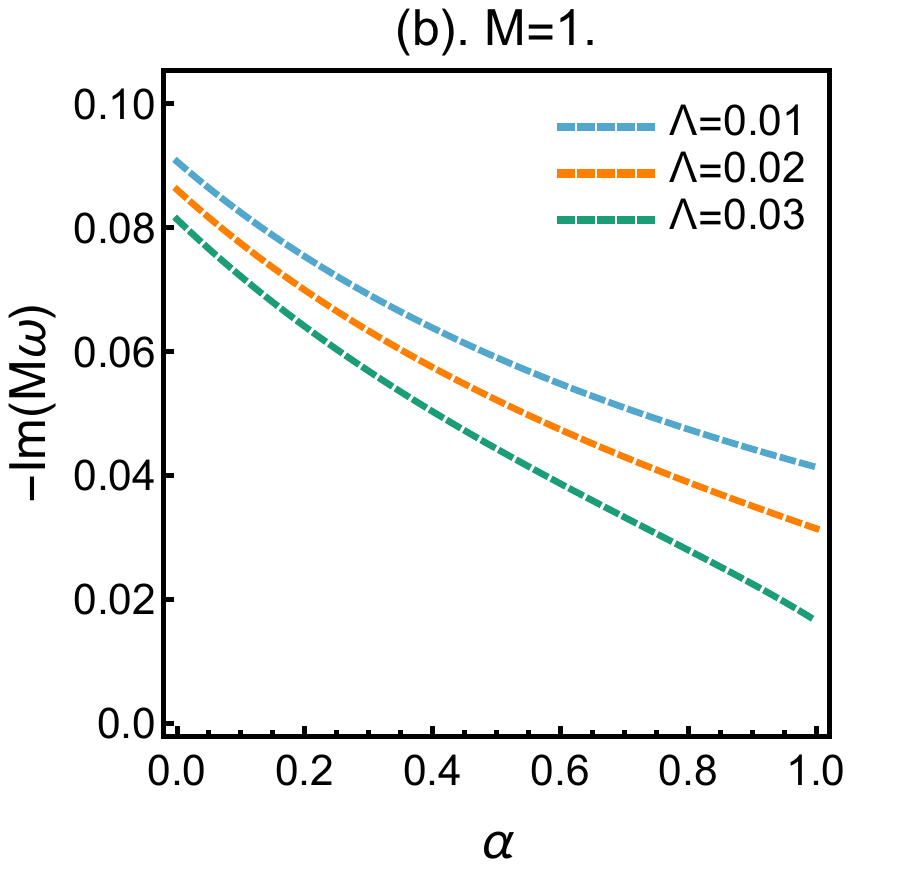}
\includegraphics[width=0.48\linewidth]{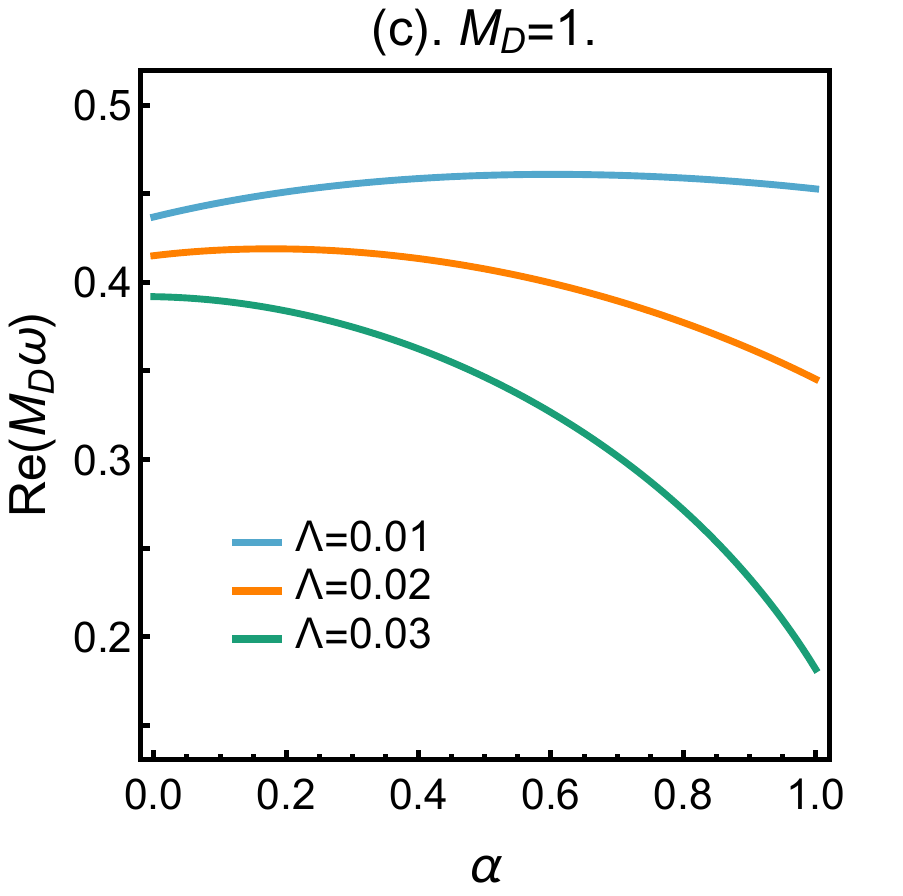}
\includegraphics[width=0.48\linewidth]{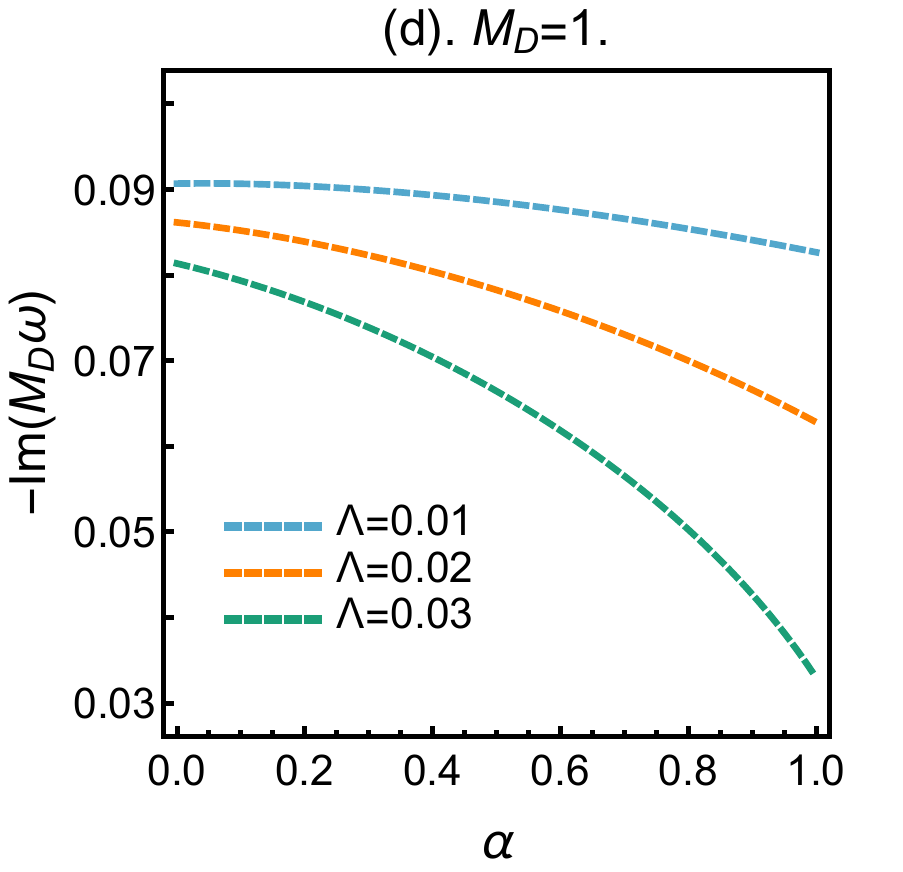}
\caption{
The axial electromagnetic frequencies in the $l=2, n=0$ mode, where (a) and (b) correspond to the results when the black hole mass $M=1$, while (c) and (d) correspond to the results when the black hole ADM mass $M_\text{D}=1$.}
\label{fig:eloddre}
\end{figure}
In Figs. \ref{fig:eloddre}-(a) and \ref{fig:eloddre}-(b), the QNMs were displayed for a scaled mass set $ M = 1 $ in order to illustrate the difference in magnitude with the result of GR.
Where take the axial electromagnetic perturbation of $l=2$ and $n=0$ as an example, the real and imaginary parts of the electromagnetic frequency are observed to rapidly decrease as the increase of parameters $\alpha $ or $ \Lambda $.
However, imposing the same scaling condition $ G M = 1 $ from GR to MOG theory, we obtain $ GM = G_\text{N}(1+\alpha)M= 1 $ thus yielding $ M_\text{D}=1 $.
Consequently, we can observe larger values of QNMs that correspond to lower mass black holes than predicted by GR.
Figs. \ref{fig:eloddre}-(c) and \ref{fig:eloddre}-(d) show these results.
\begin{figure}[h]
\centering
\includegraphics[width=0.48\linewidth]{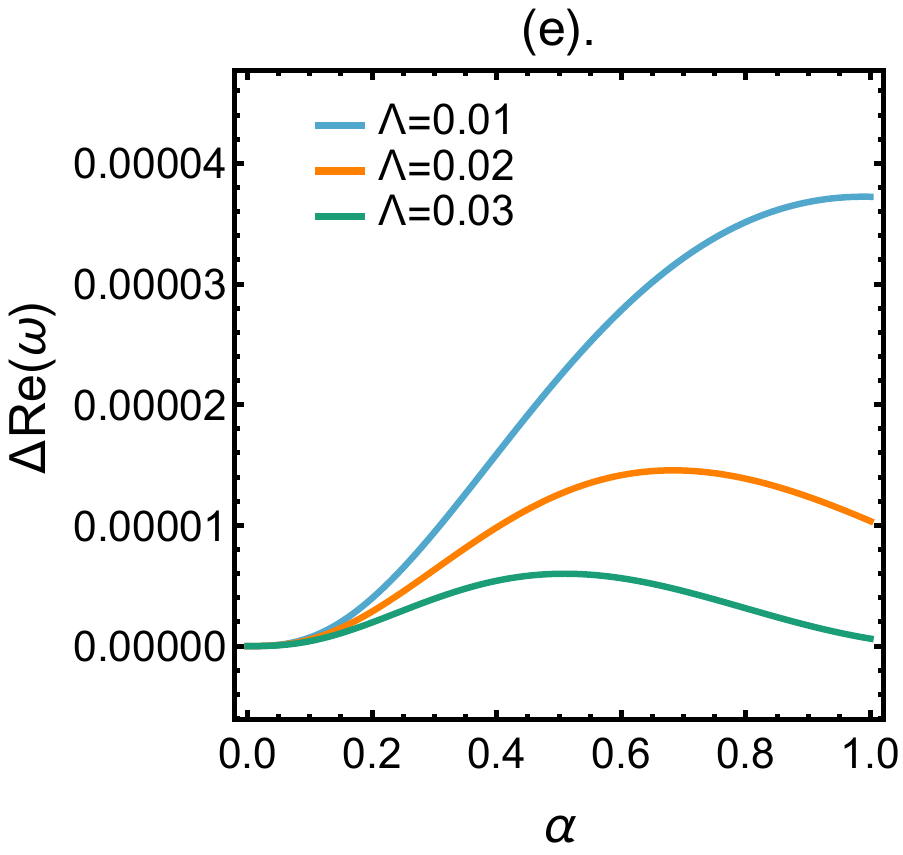}
\includegraphics[width=0.48\linewidth]{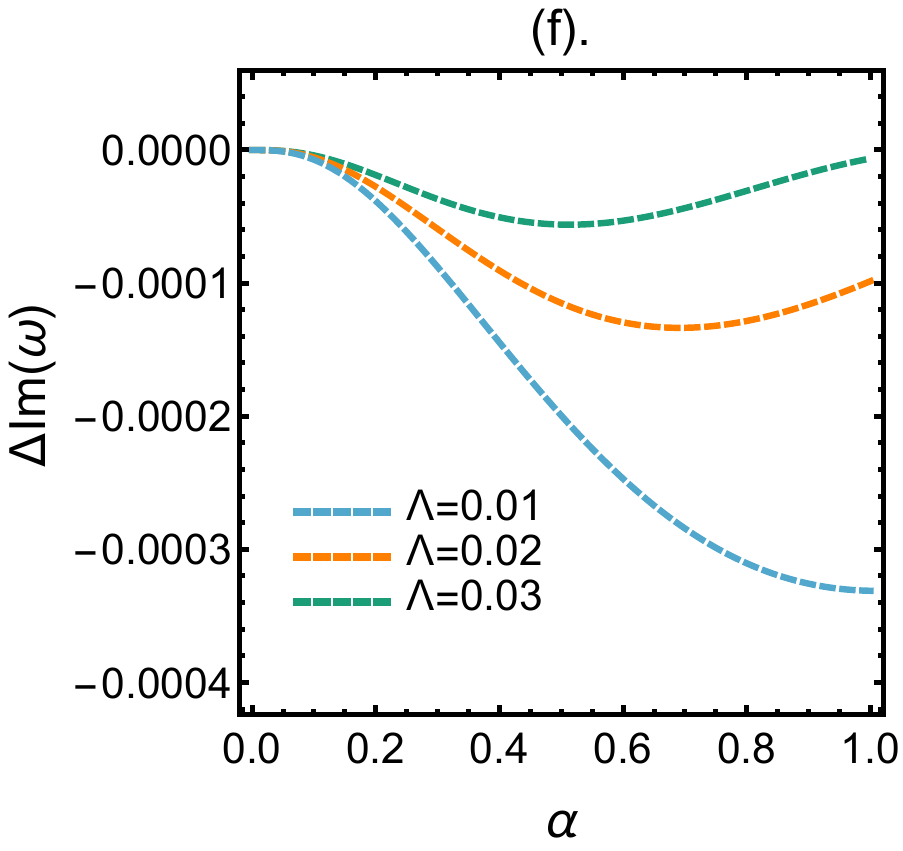}
\includegraphics[width=0.48\linewidth]{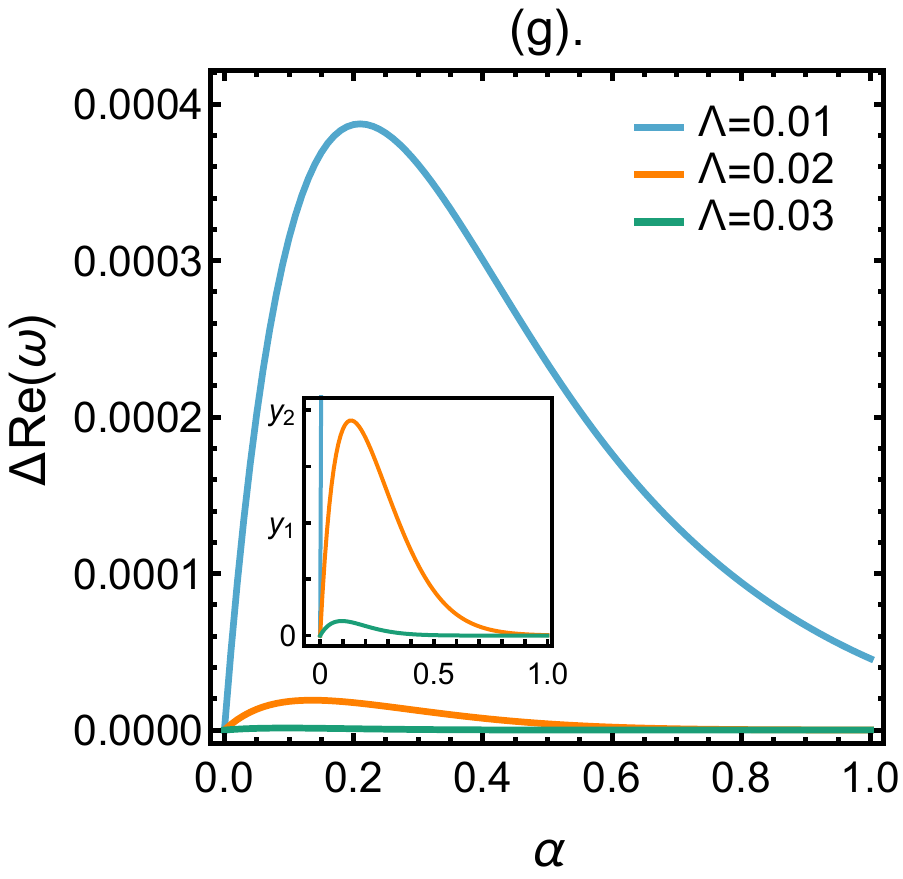}
\includegraphics[width=0.48\linewidth]{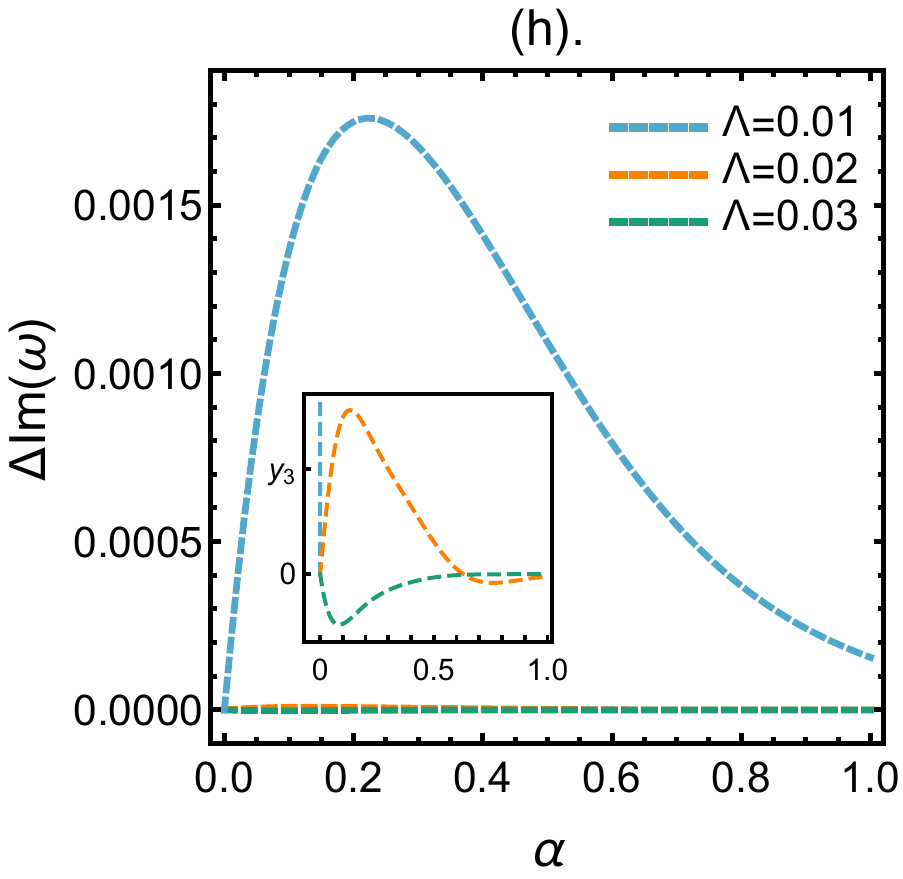}
\caption{The percentage error between its corresponding axial and polar sections in the $ l=2, n=0 $ electromagnetic mode.
The results of the WKB approach are shown at the (e) and (f), while the results of the matrix method are displayed at the (g) and (h).
Note, the coordinates $ \text{y}_1=10^{-5} $, $ \text{y}_2=2\times10^{-5} $ and $ \text{y}_{3}=5\times 10^{-6} $.
}
\label{fig:errorEM}
\end{figure}
In Appendix \ref{AppendixQNM}, one can easily determine from the data that the electromagnetic and gravitational mode frequencies display strict isospectrality in the axial and polar sections.
Therefore, we can perform an error analysis of the isospectral properties to assess the accuracy of the numerical method.
Figs. \ref{fig:errorEM}-(e) and \ref{fig:errorEM}-(f) show the percentage error between the different parities under the WKB approach, while Figs. \ref{fig:errorEM}-(g) and \ref{fig:errorEM}-(h) correspond to the matrix method.
Here, the percentage error of the  real part, for example, is defined as
\begin{equation}
\Delta \text{Re}\left(\omega\right)=100\times\frac{\text{Re}(\omega_\text{po})-\text{Re}(\omega_\text{ax})}{\text{Re}(\omega_\text{ax})}.
\end{equation}
And it can be observed that the errors generated by both methods converge as the cosmological constant $\Lambda$ is increased. Specifically, the error of the matrix method exhibits exponential convergence.
Additionally, it is worth noting that the WKB approach exhibits higher accuracy when both the MOG parameter $ \alpha $ and cosmological constant $\Lambda$ are small, with values such that $\alpha<0.1$ and $\Lambda<0.01$.

\subsubsection{The gravitational modes}\label{Gramode}
The gravitational modes exist only for $ l\geq 2 $. During the merger of black hole, $l=2$ mode will provide significant contribution.
According to the findings of these reported in Refs. \cite{Cardoso:2016olt,Wei:2018aft,Brito:2018hjh}, the modes under consideration may play a significant role in the radiation emitted during the merger of black holes.
As predicted by these studies, it is a generic characteristic that the process in question is accompanied by the simultaneous emission of both electromagnetic and gravitational waves. The ringdown phase is characterized by a superposition of both electromagnetic and gravitational QNM frequencies.
\begin{figure}[h]
\centering
\includegraphics[width=0.48\linewidth]{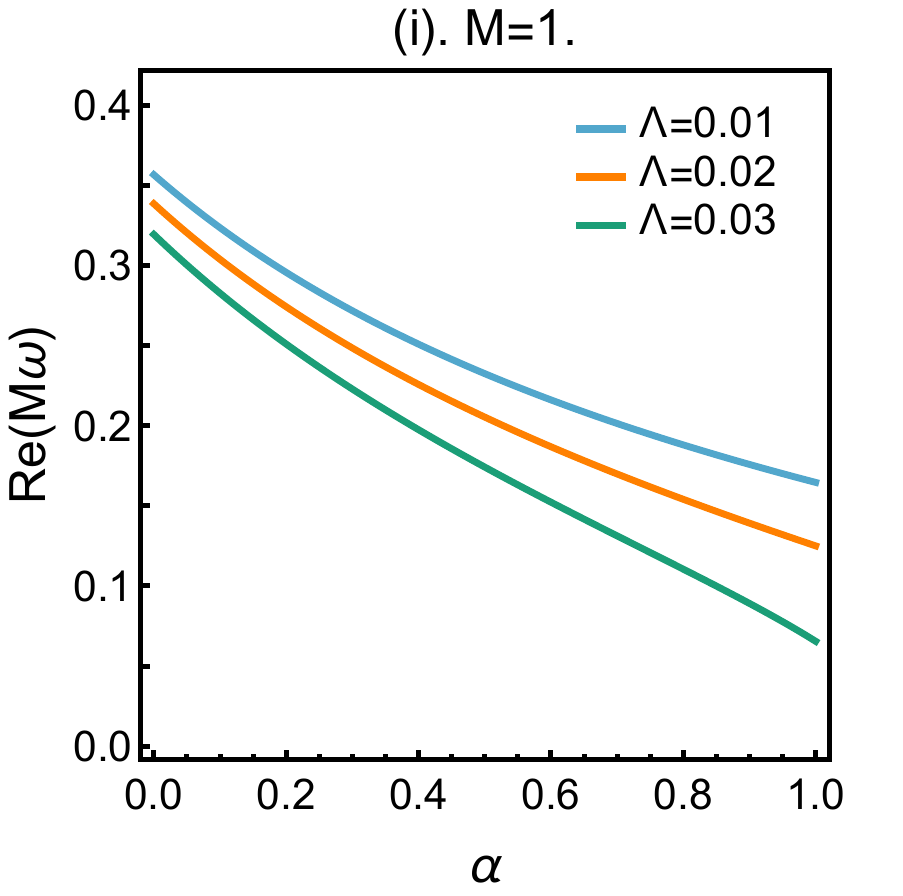}
\includegraphics[width=0.48\linewidth]{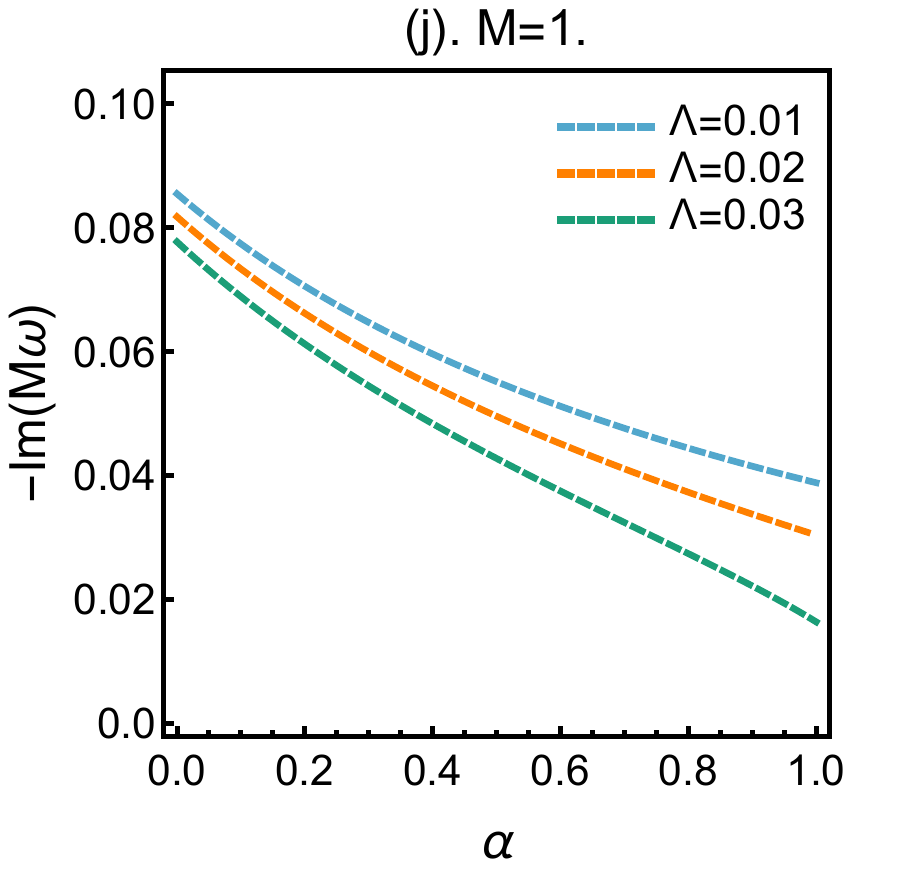}
\includegraphics[width=0.48\linewidth]{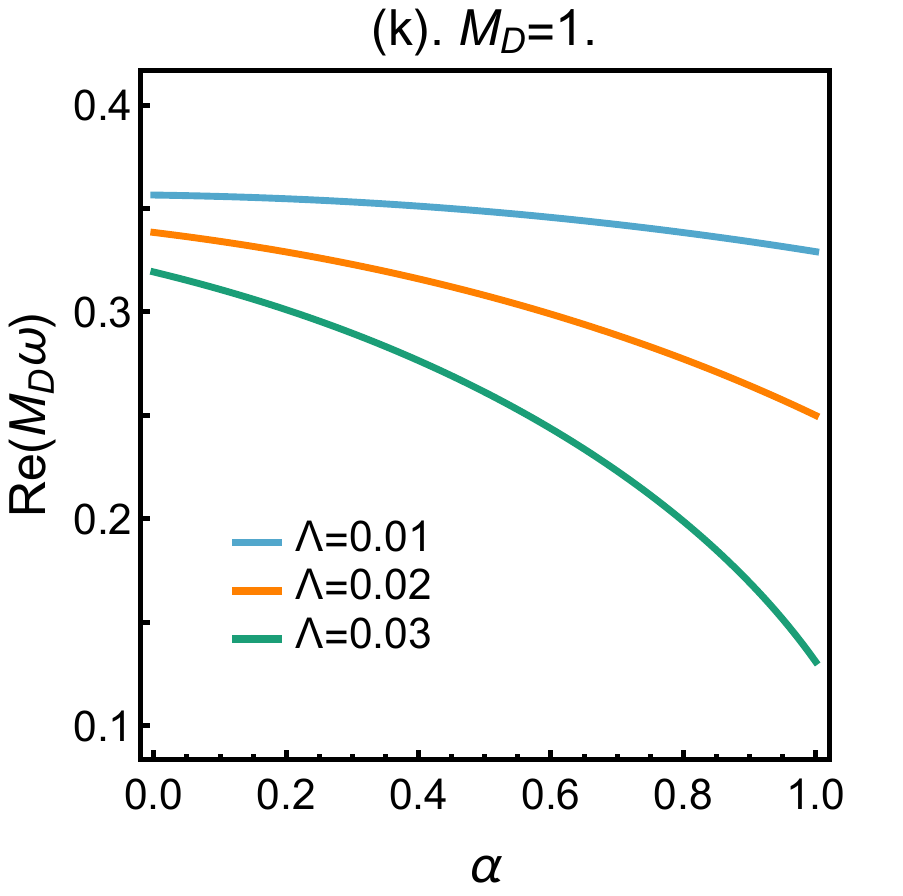}
\includegraphics[width=0.48\linewidth]{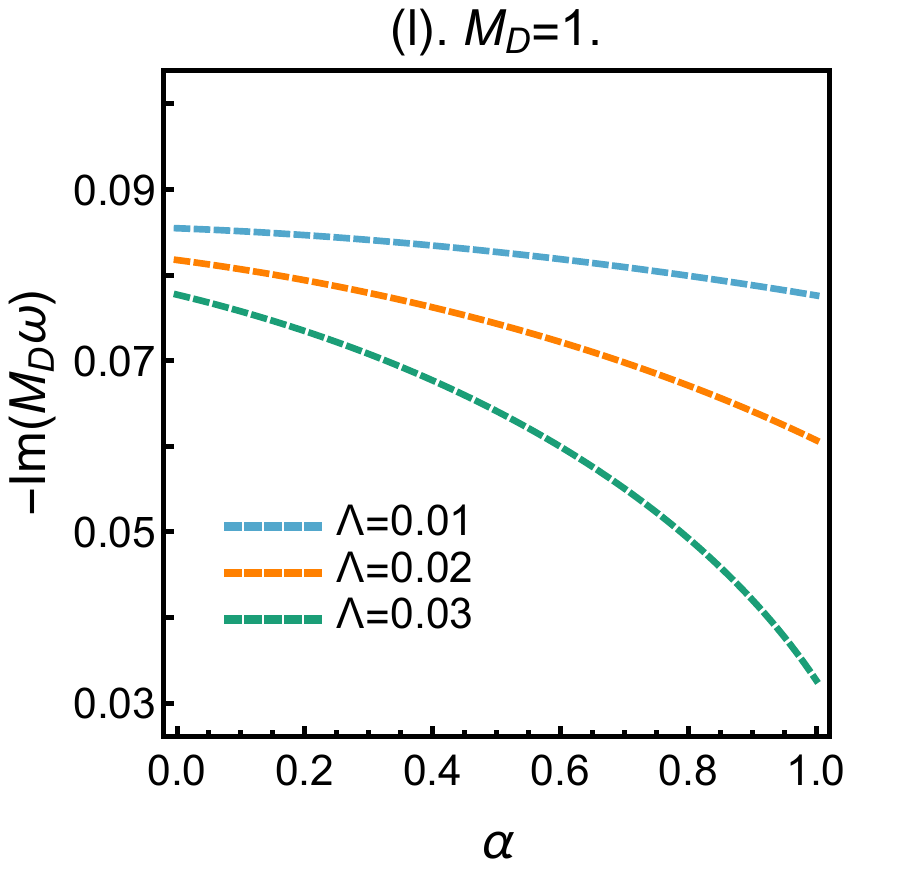}
\caption{The axial gravitational frequencies in the $l=2, n=0$ mode, where (i) and (j) correspond to the results when the black hole mass $M=1$, while (k) and (l) correspond to the results when the black hole ADM mass $M_\text{D}=1$.}
\label{fig:groddre}
\end{figure}
Figs. \ref{fig:groddre} shows that, compared with the electromagnetic QNMs, the gravitational frequencies have a similar trend in response to parameter $\alpha$ as the electromagnetic frequency, however, with a lower magnitude and a smoother decay rate.
In addition, comparison with Figs. \ref{fig:eloddre}, as the parameter $\alpha$ varies, the gravitational frequencies appear to be less affected than the electromagnetic perturbation.
\begin{figure}[h]
\centering
\includegraphics[width=0.48\linewidth]{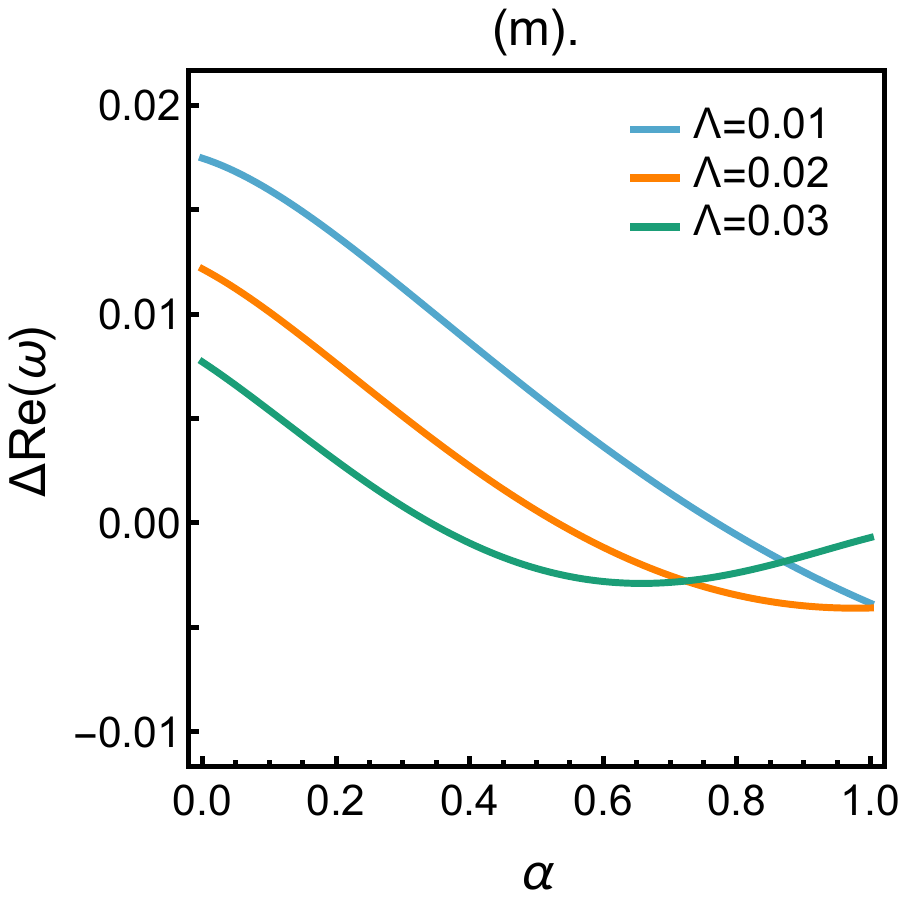}
\includegraphics[width=0.48\linewidth]{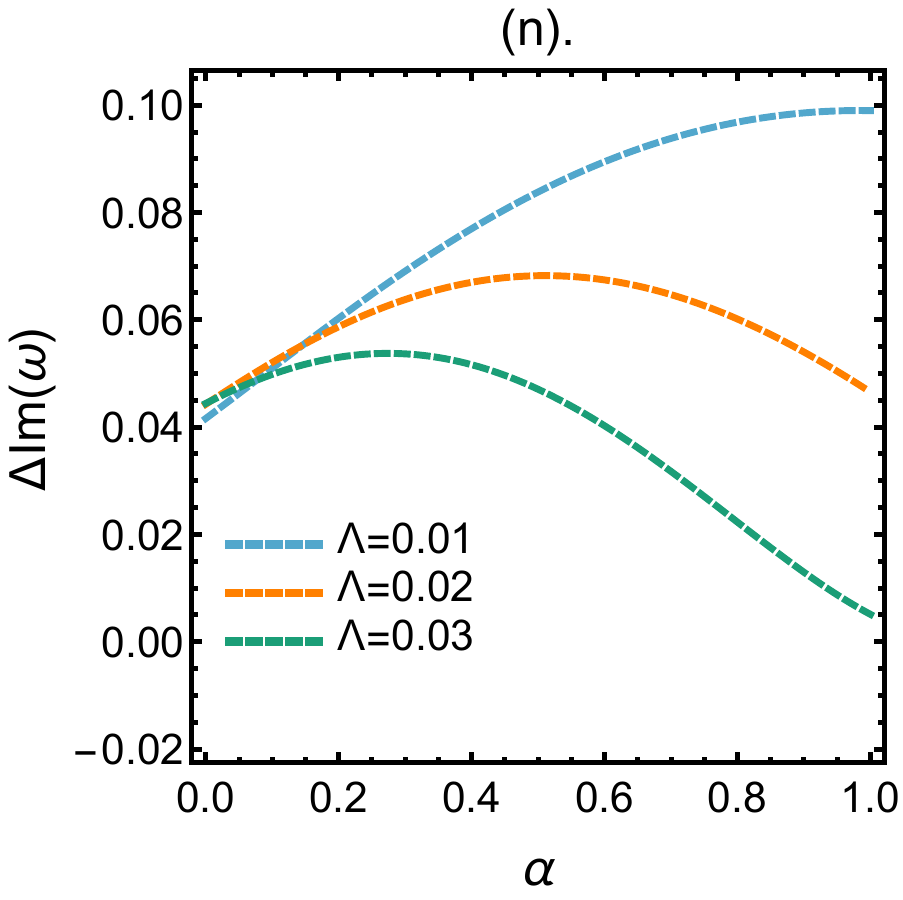}
\includegraphics[width=0.48\linewidth]{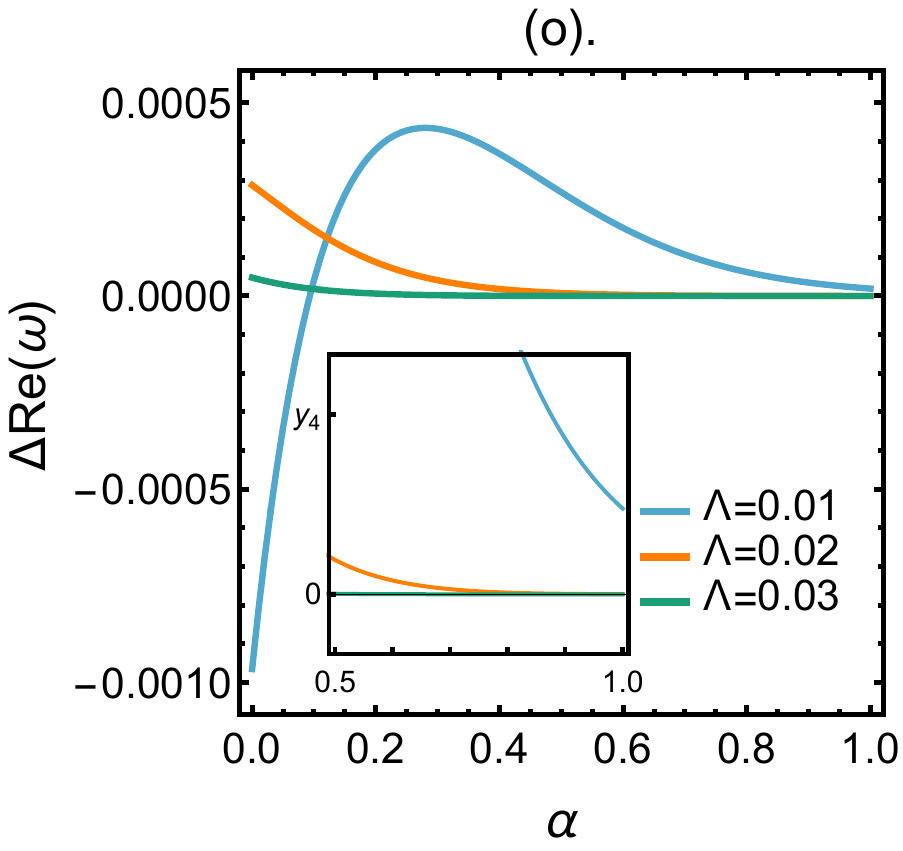}
\includegraphics[width=0.48\linewidth]{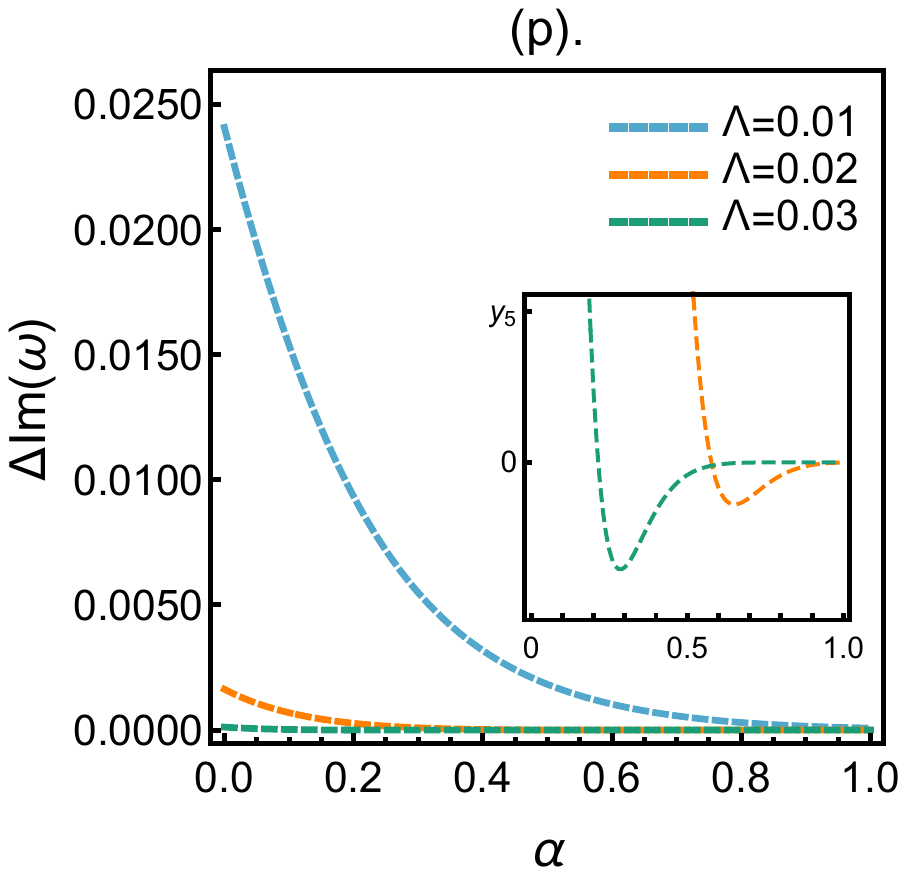}
\caption{The percentage error between its corresponding axial and polar sections in the $ l=2, n=0 $ gravitational mode.
The results of the WKB approach are shown at the (m) and (n), while the results of the matrix method are displayed at the (o) and (p).
Note, the coordinates $ \text{y}_4=4\times10^{-5} $ and $ \text{y}_{5}=2\times 10^{-6} $.
}
\label{fig:grerror}
\end{figure}
In Fig. \ref{fig:grerror}, we can observe that the WKB approach introduces an error of $ 0.1\% $.
Note, the generation of large errors depends more on the WKB approach, since the precision of this method is related to the fields being considered \cite{Zhidenko:2003wq}.
At the same time, the matrix method still maintains highly accurate numerical results in gravitational modes.
Similarly to electromagnetic modes, in Figs. \ref{fig:grerror}-(o) and \ref{fig:grerror}-(p), the errors decrease with increasing of the cosmological constant $\Lambda$, as determined by the property of Eq. \meq{psi} in the matrix method.
Consequently, we consider that the matrix method is expected to yield better accuracy for gravitational frequencies.

It is well-known that the gravitational and the electromagnetic perturbations of the Schwarzschild and the RN black holes exhibit an important property called isospectrality, which was first demonstrated by Chandrasekhar \cite{Chandrasekhar:1984siy}.
Despite the fact that the axial and polar sectors of the perturbations are governed by distinct potentials, their QNM spectra remain identical \cite{Berti:2009kk}.
However, this isospectral property does not apply to modified theories, such as loop quantum gravity \cite{isoLoop}, Chern-Simons gravity \cite{isoChSi} and Lovelock gravity \cite{isoLovelock}, or in higher-dimensional spacetimes \cite{Berti:2009kk}.
Noteworthy, in the case of AdS spacetimes with a negative cosmological constant, isospectrality breaking occurs in the most common S-AdS spacetimes.
This implies that isospectrality in the MOG-AdS spacetimes, which is a generalization of the S-AdS spacetimes, should also be broken due to their particular boundary conditions.
Therefore, the focus of our study is isospectrality in MOG-dS spacetimes, and based on the data presented in Appendix \ref{AppendixQNM} and Figs. \ref{fig:errorEM}, \ref{fig:grerror},
we confirm that isospectrality is indeed present in this scenario.

\subsection{Case for $ \xi=1 $}

In this subsection, we will explore the influence of the interaction term on the QNM spectrum; for convenience, we set $\Lambda=0$.
For equations governing QNMs that are coupled, the matrix method and the WKB approach are no longer applicable. 
Instead, we employ the matrix-valued continued fraction method.
Following Leaver's foundational research \cite{Leaver1985}, it is a well-established fact that continued fraction techniques can resolve the eigenvalue problems in GR. 
Specializing in Schr\"{o}dinger-like potentials, this method proves to be particularly effective when dealing only with (fractions of) terms that are powers of $ 1/r $. 
In this method, the eigenfunctions can be written as series in which the coefficients satisfy finite-term recurrence relations.
Subsequently, the continued fraction method was generalized to solve coupled system of equations, such as \eqs{odd}; further discussion can be found in Ref. \cite{Pani:2013pma}.

The coupled \eqs{odd} can be rewritten into a compact form
\begin{align}\label{odd1}
\frac{d^2}{dr^2_*}\mathbf{Y}+(\omega^2-\mathbf{V})\mathbf{Y}=0,
\end{align}
where 
\begin{align}
\mathbf{Y}=\left(
\begin{array}{c}
 \psi_g\\
\psi_e
\end{array}
\right)
\end{align}
and $ \mathbf{V} $ is a $ 2\times 2 $ matrix, ie.,
\begin{align}
\mathbf{V}=\left(\begin{array}{cc}
\frac{F(2-\lambda+rF')}{r^2}-\frac{2F^2}{r^2} & \frac{4i\omega M \sqrt{\alpha}(1+\alpha)F}{r^{7/2}\lambda} \\
-\frac{iM\sqrt{\alpha}\lambda \mathcal{D}_2}{r^{7/2}\omega} & -\frac{F(4\lambda+7F-2rF')}{4r^2}
\end{array}\right).
\end{align}
In order to reduce the equation to a matrix-valued recurrence relation, we use the ansatz
\begin{equation}
\begin{aligned}
\psi_g=&\frac{r_h(r_h-r_m)^{-4(1+\alpha)iM\omega -1}}{r}e^{-2i\omega r_h}e^{i\omega r}\\
&\times(r-r_m)^{1+2(1+\alpha)iM\omega}z^{\frac{-i\omega r^2_m}{r_h-r_m}}a^{(1)}_nz^n 
\end{aligned}
\end{equation}
and
\begin{equation}
\begin{aligned}
\psi_e=&\frac{r_h(r_h-r_m)^{-4(1+\alpha)iM\omega -1}}{\sqrt{r}}e^{-2i\omega r_h}e^{i\omega r}\\
&\times(r-r_m)^{1+2(1+\alpha)iM\omega}z^{\frac{-i\omega r^2_m}{r_h-r_m}}a^{(2)}_nz^n ,
\end{aligned}
\end{equation}
where $ z=(r-r_h)/(r-r_m) $ .
In this case, we obtain a 6-term matrix-valued recurrence relation for the vectors $ \mathbf{a}_n $.
It can be reduced to a three-term recurrence relation using a matrix analog of Gaussian elimination \cite{Pani:2013pma}.
Here, we verified the results for different values of $ n $, which showed clear convergence.
As mentioned in Sec. \ref{QNMs0}, the electromagnetic modes exist for $ l\geq 1 $.
However, this does not apply to the current situation, as we need to solve both electromagnetic and gravitational modes simultaneously, hence $ l\geq 2 $. In the following, we focus on the modes for $l=2$ and $l=3$.

\begin{figure}[h]
\centering
\includegraphics[width=0.47\linewidth]{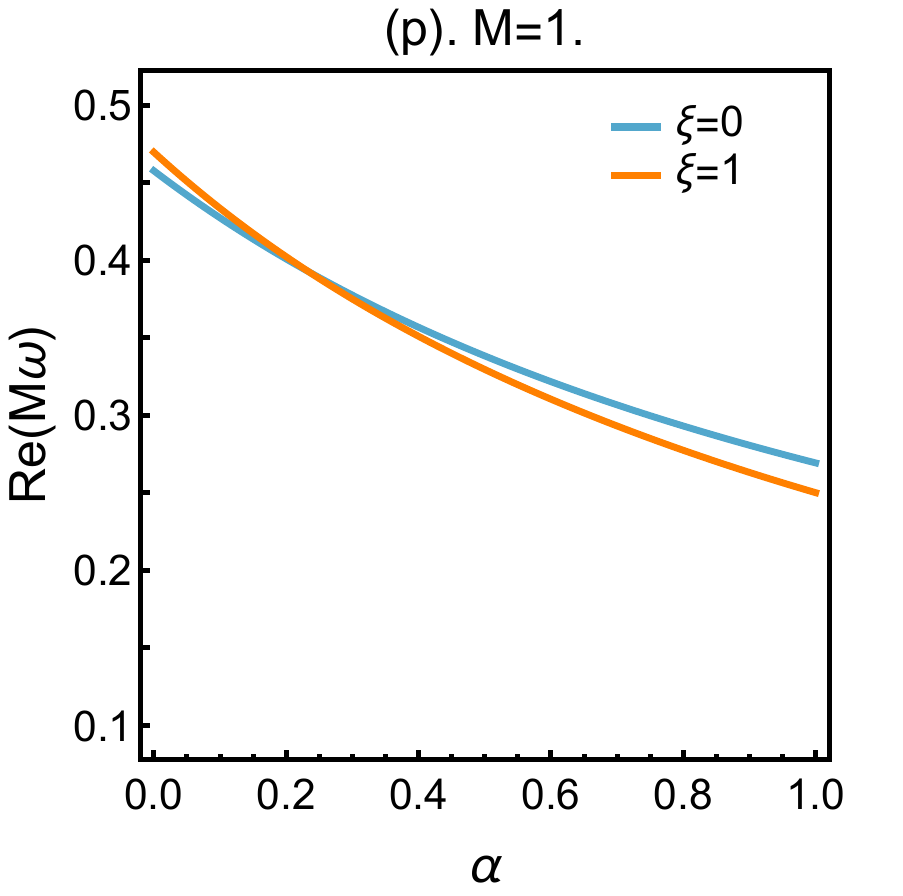}
\includegraphics[width=0.48\linewidth]{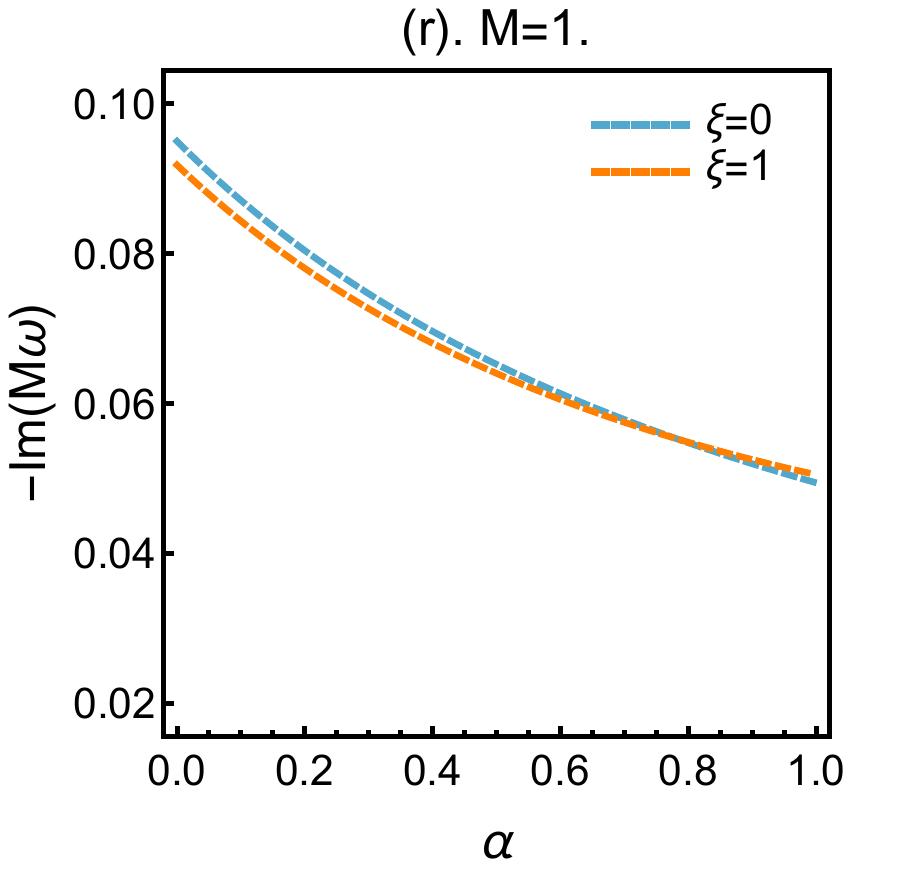}
\includegraphics[width=0.47\linewidth]{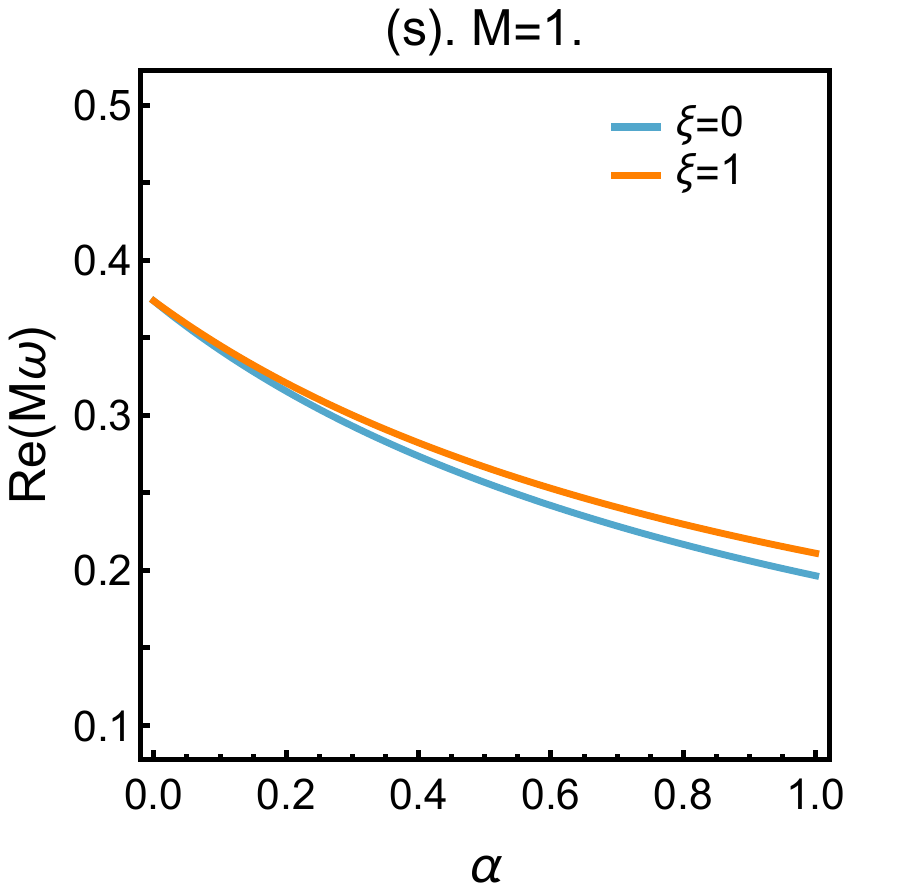}
\includegraphics[width=0.48\linewidth]{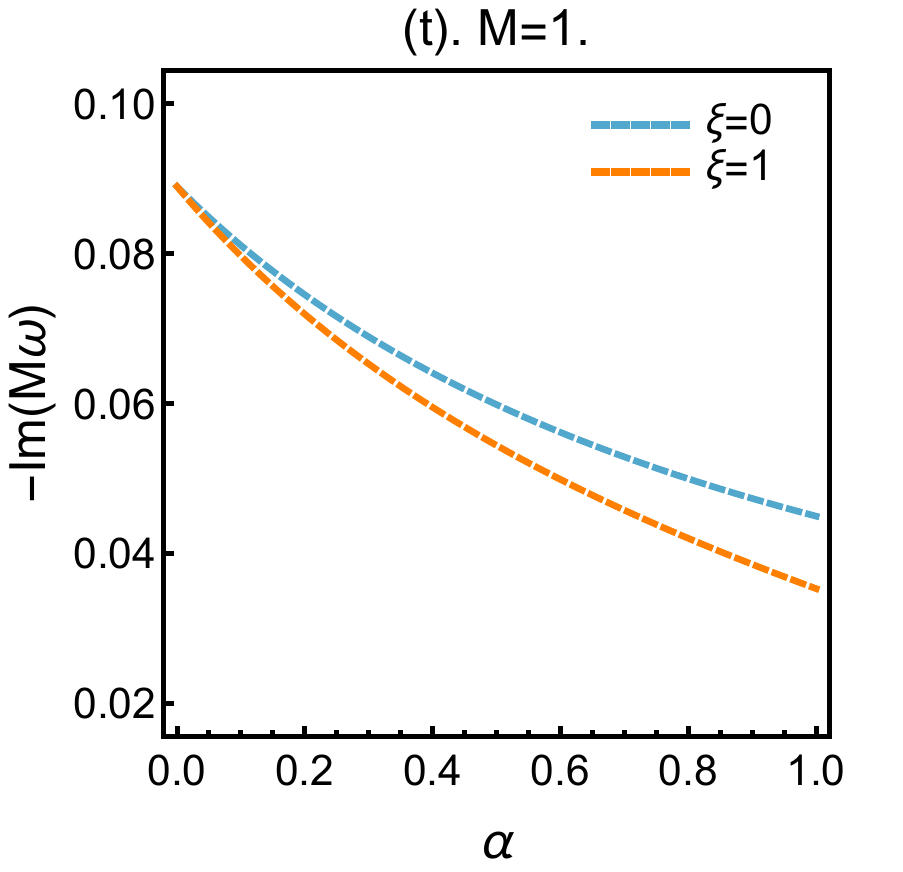}
\caption{In the $ l=2, n=0 $ mode, the figures (p) and (r) represent electromagnetic frequencies, while (s) and (t) represent gravitational frequencies.}
\label{xi1}
\end{figure}
In MOG theory, the effects of the interaction term on the gravito-electromagnetic perturbations in black holes are depicted in Figs. \ref{xi1}.
When $ \alpha=0 $, the electromagnetric and the gravitational modes will naturally decouple. 
This results in the interaction term solely influencing the electromagnetic mode, but leaving the gravitational mode unaffected. 
The introduction of the MOG parameter $\alpha$ will cause the coupling between these two modes, and lead to a noticeable deviations.
Specifically, when we considering the interaction term, as the MOG parameter increases, the real part of the gravitational mode decreases at a slower rate, whereas the imaginary part decreases more rapidly.
While for the electromagnetic modes, as the increase of $\alpha$, considering the interaction term will affect the QNM frequencies as shown in Figs. \ref{xi1}. When $l=3$, the results is similar to the case $l=2$.

\begin{figure}[h]
\centering
\includegraphics[width=0.8\linewidth]{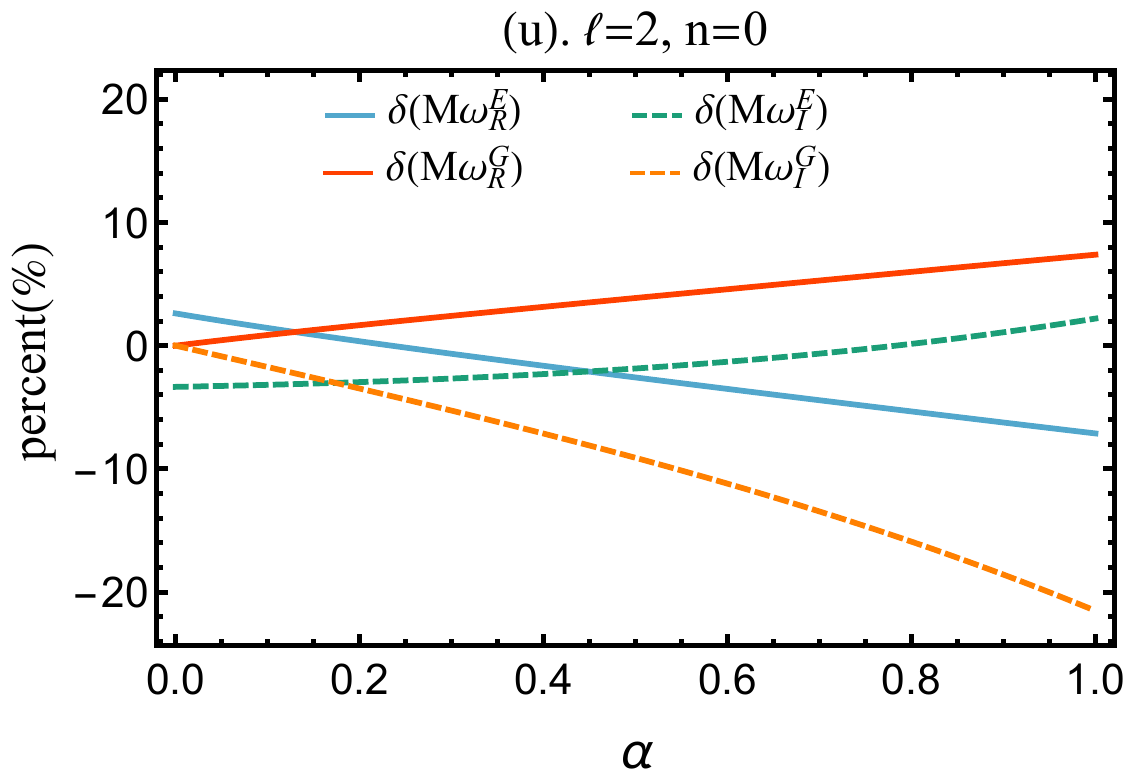}
\includegraphics[width=0.8\linewidth]{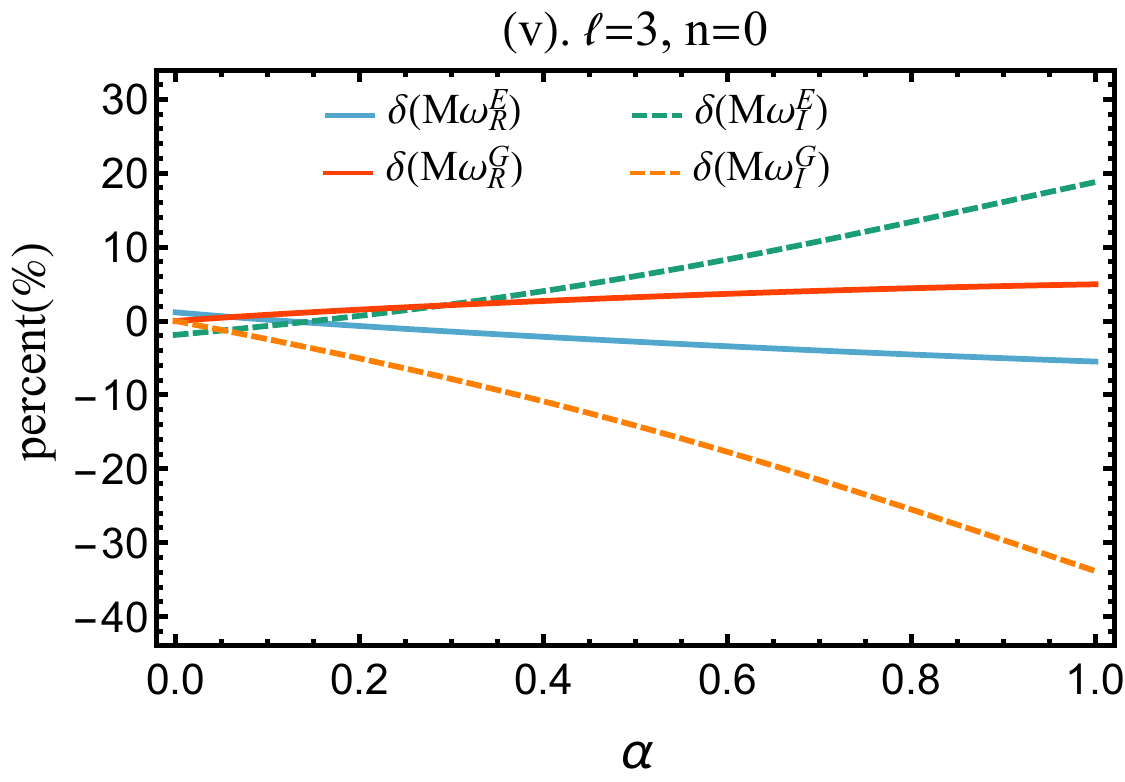}
\caption{The percentage deviation introduced by the interaction terms in the $ l=2 $ and $ l=3 $ modes varies with the MOG parameter.}
\label{xi2}
\end{figure}
In Figs. \ref{xi2}, we show the percentage deviation of the modes $ l = 2 $ or $ l = 3  $, which is defined as
\begin{equation}
\begin{aligned}
\delta\left(M\omega\right)=& 100\times \frac{M\omega_{(\xi=1)}-M\omega_{(\xi=0)}}{M\omega_{(\xi=0)}}.
\end{aligned}
\end{equation}
Our results show that, for small $\alpha$, higher value of $l$ results smaller devation, since the term that dominates the frequency is determined by the $F\lambda/r^2$ term in the potential function. 
However, when the MOG parameter $\alpha$ is large enough, the effect of the interaction term becomes non-negligible. The reason is that the source term contains the coefficient $\lambda$, which leads to an increasing deviation as $ l $ increases.

\section{Ringdown Waveforms}\label{Sec.5}
To investigate the contribution of all modes of the electromagnetic and gravitational  perturbation of the MOG-dS black holes, one could perform numerical simulations to solve the perturbation equations for the black holes.
In a finite time domain, we can consider the numerical evolution of an initial wave packet governed by the time-dependent Schr\"{o}dinger-like equation. Expressing Eq. \meq{mastereq} as
\begin{align}\label{mastereqt}
\frac{\pp^2}{\pp r_*^2}\psi-\frac{\pp^2}{\pp t^2}\psi-V\psi=0,
\end{align}
and using the light-cone coordinates $ u = t-r_* $ and $ v = t + r_* $ \cite{Gundlach:1993tp}, the above equation can be written as
\begin{align}\label{masterequv0}
4\frac{\pp^2 \psi(u,v)}{\pp u\pp v}-V(u,v)\psi(u,v)=0.
\end{align}
The use of light-cone coordinates can simplify the analysis of gravitational waves, since the equation takes on a simple form in these coordinates.
In particular, the equation becomes a wave equation in $u$ and $v$, which can be solved using standard techniques such as finite difference method (FDM) \cite{Abdalla:2010nq,Zhu:2014sya,Lin:2022owb,Fu:2022cul,Tan:2022vfe}.
We compute the waveform $ \psi(u,v) $ by imposing the following initial conditions for Eq. \meq{masterequv0}
\begin{align}\label{masterequv}
\psi(u,0)=0,~~\psi(0,v)=\text{Exp}\left(-\frac{(v-v_c)^2}{2\gamma^2}\right),
\end{align}
where $ \psi(0,v) $ is a Gaussian wave packet centered at $v_c$ and having a width of $\gamma$.
The observer position, located at $r_0=10 r_h$ with Boyer-Lindquist coordinates, is situated in the outer communication domain and satisfies the condition $r_h < r_0 < r_c$.
Then, we can numerically solve the partial differential Eq. \meq{masterequv} to generate the ringdown waveforms.
Note that our goal is to perform a time evolution and extract the time-domain waveform $ \psi(t) $ at future null infinity.
\begin{figure}[h]
\centering
\includegraphics[width=0.47\linewidth]{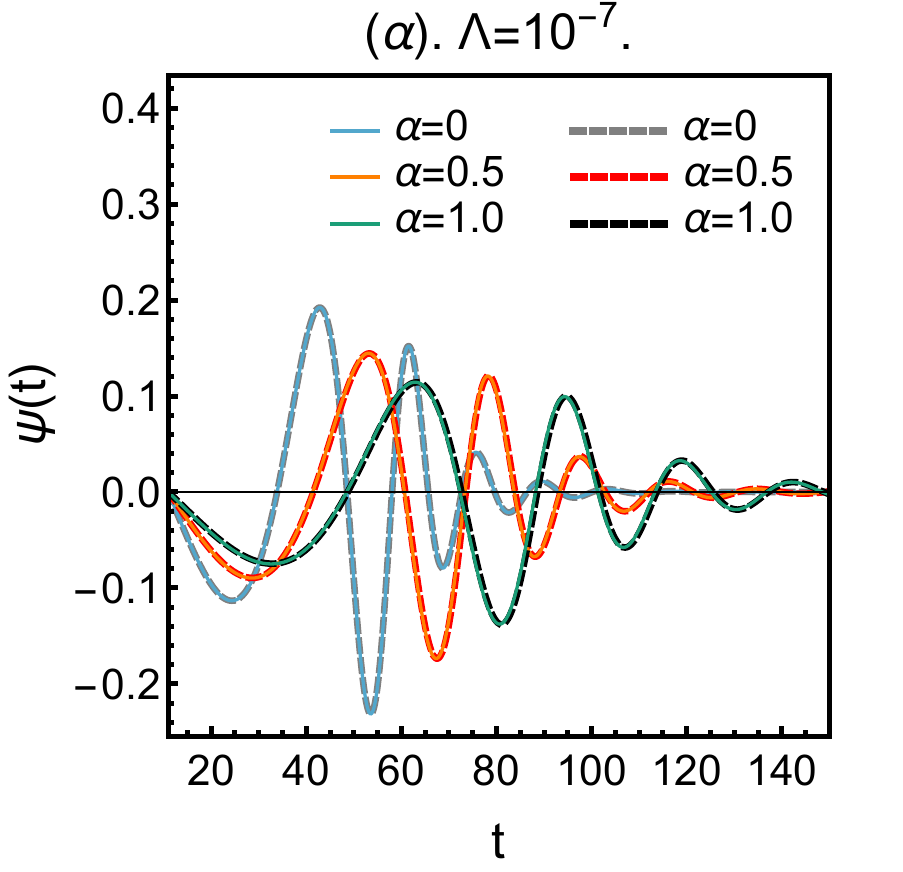}
\includegraphics[width=0.48\linewidth]{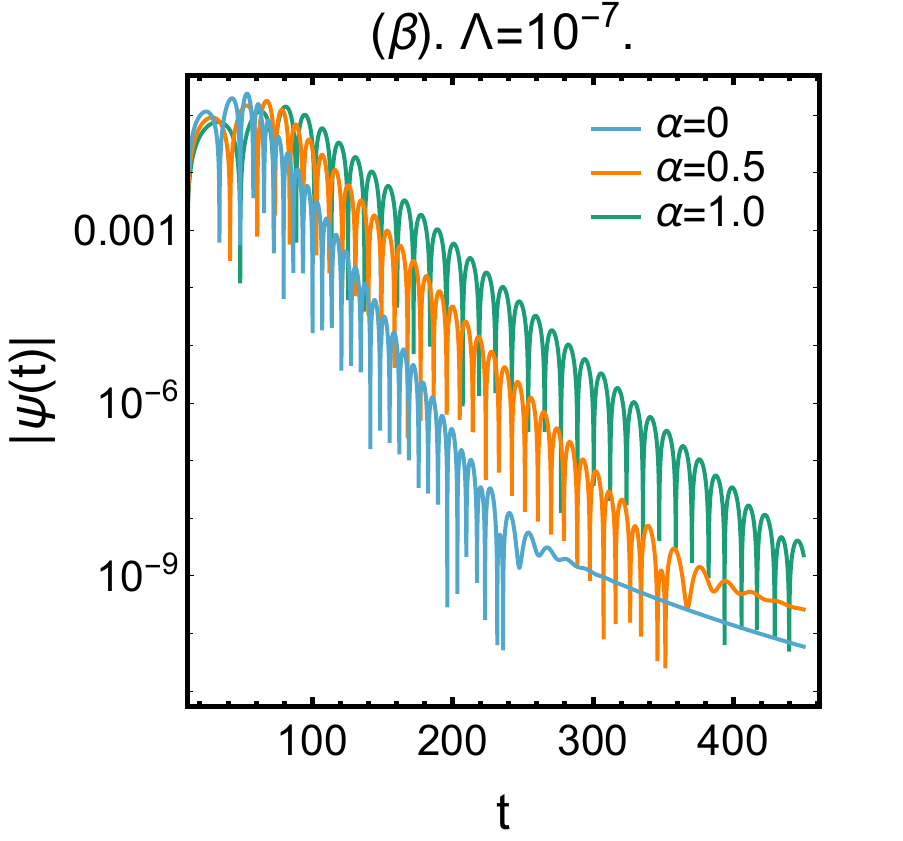}
\includegraphics[width=0.47\linewidth]{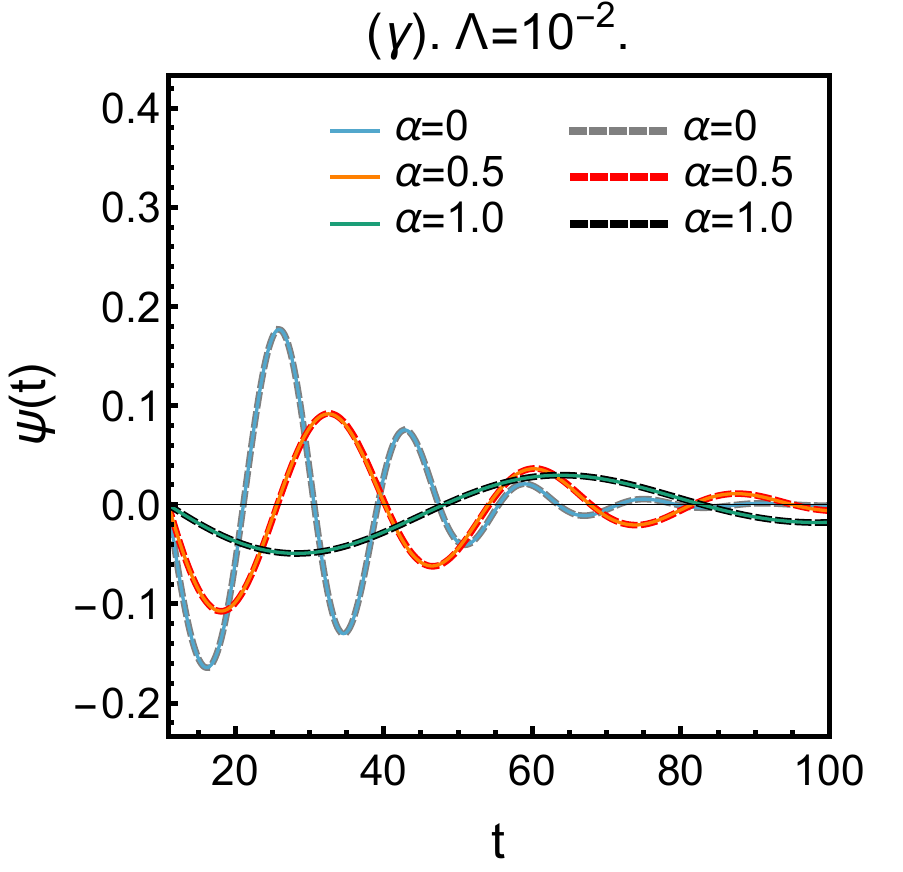}
\includegraphics[width=0.48\linewidth]{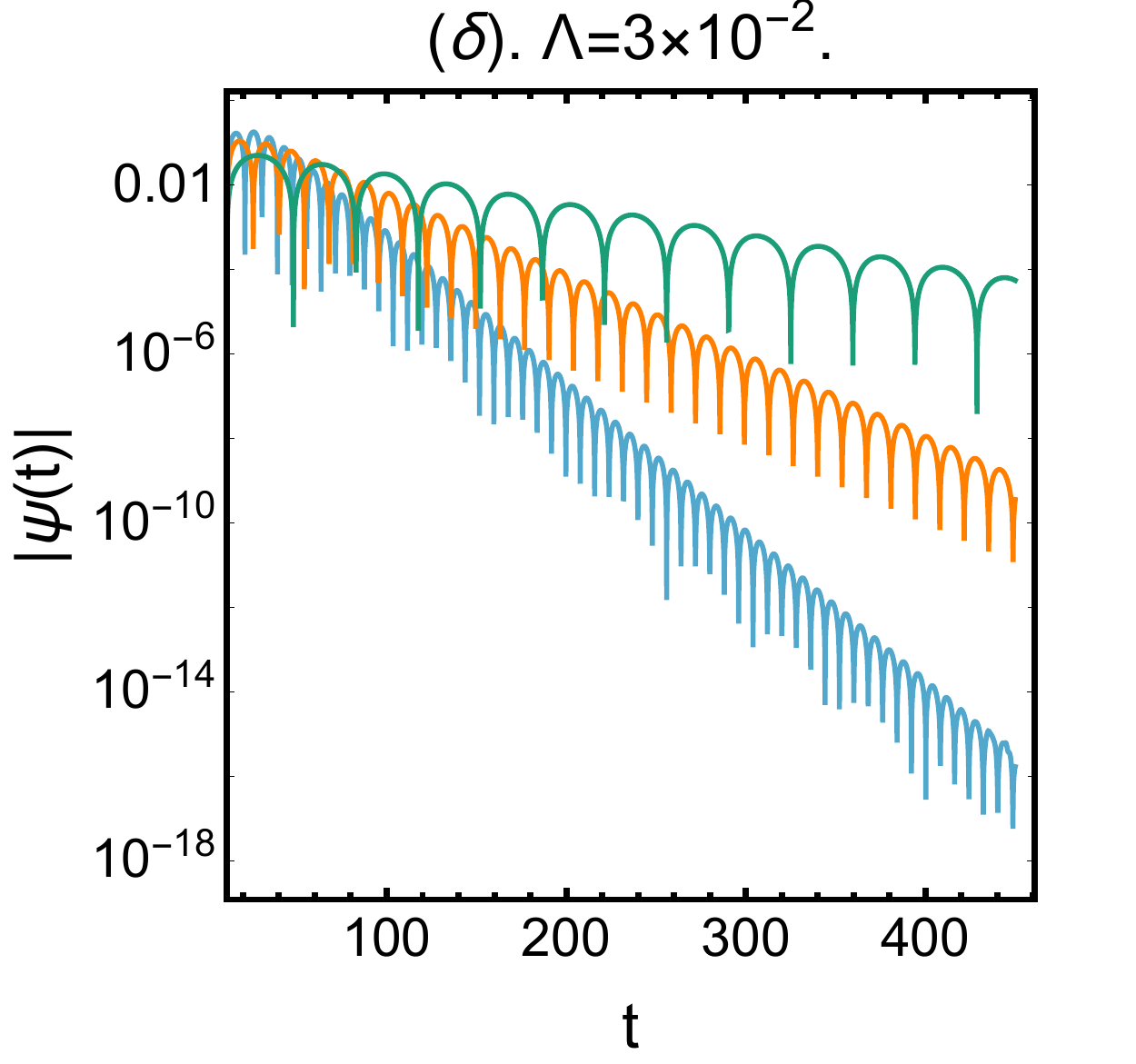}
\caption{The time evolution of the wave function $ \psi(t) $ corresponds to the axial (dotted line) and polar (full line)  electromagnetic perturbations in the $ l=2, n=0 $ mode.}
\label{EMwaves}
\end{figure}
\begin{figure}[h]
\centering
\includegraphics[width=0.47\linewidth]{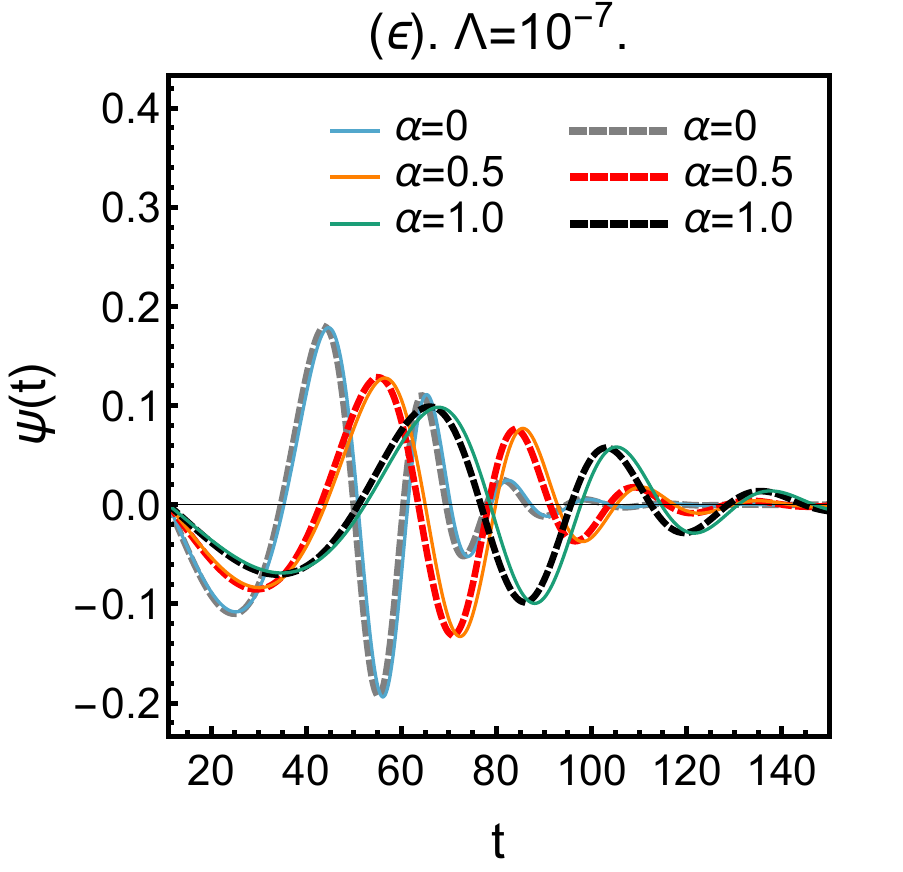}
\includegraphics[width=0.48\linewidth]{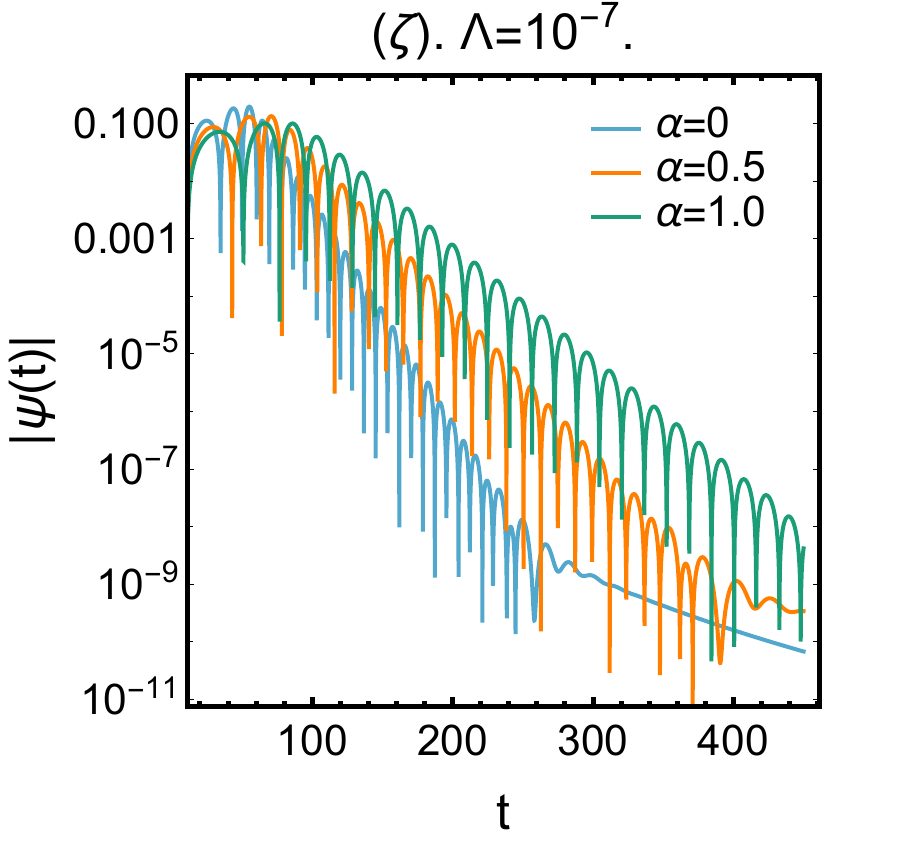}
\includegraphics[width=0.47\linewidth]{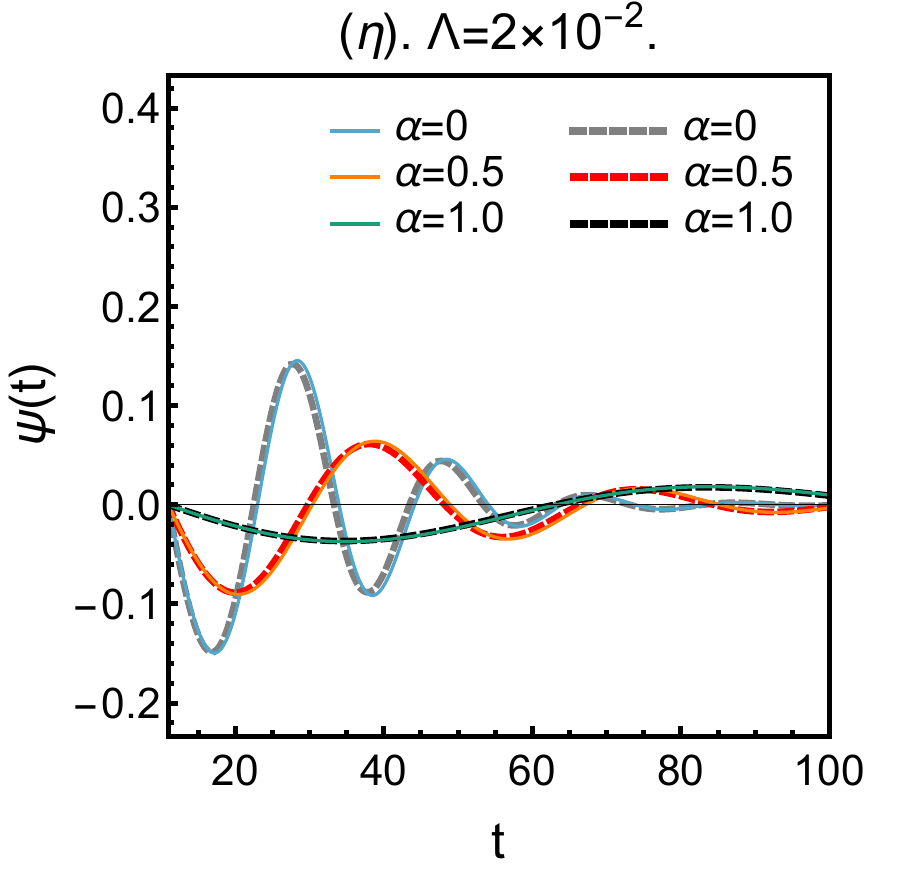}
\includegraphics[width=0.48\linewidth]{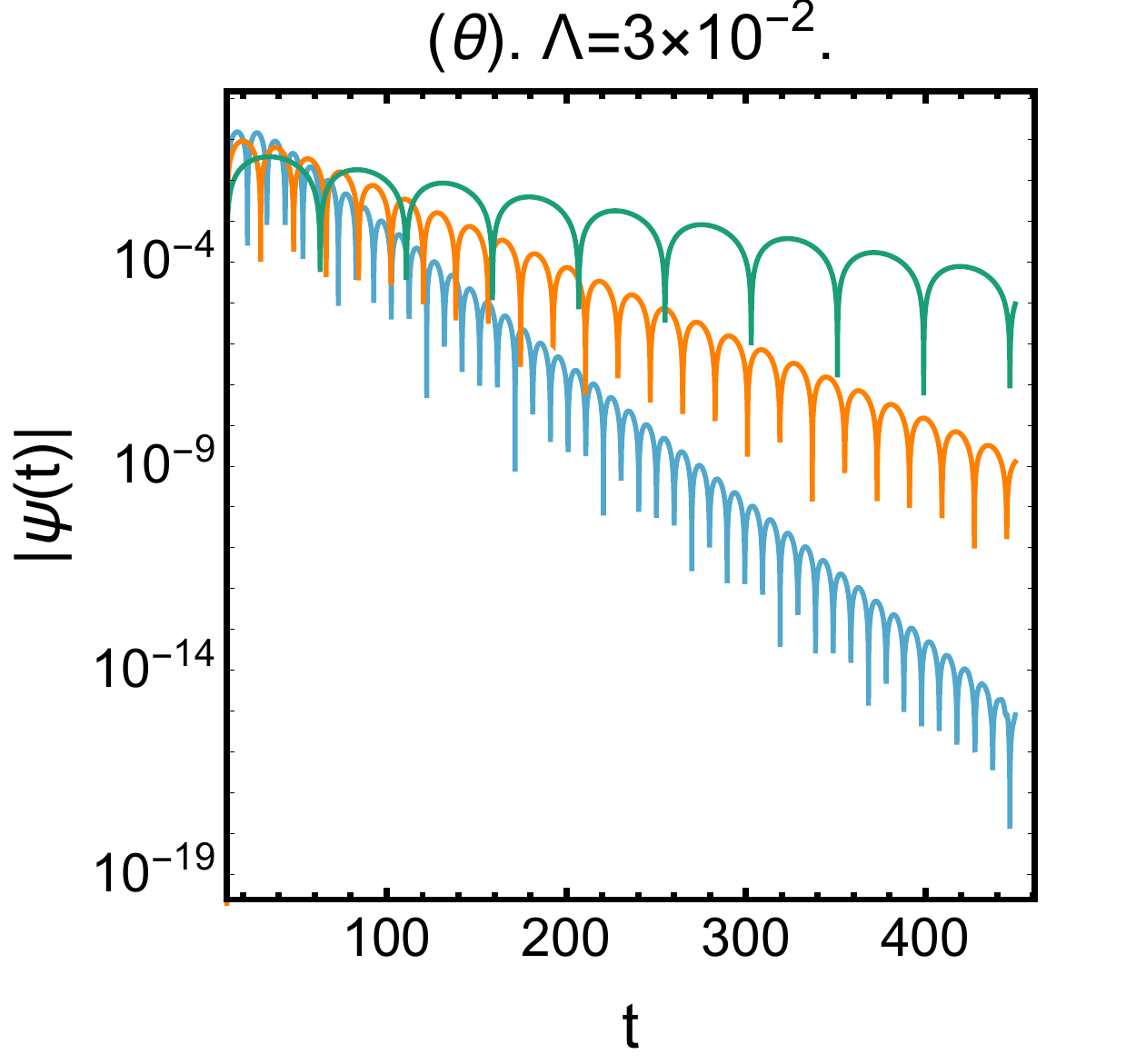}
\caption{The time evolution of the wave function $ \psi(t) $ of the axial (dotted line) and polar (full line)  gravitational perturbation in the $ l=2, n=0 $ mode.}
\label{Gwaves}
\end{figure}

The ringdown waveforms of the electromagnetic and gravitational modes at $l=2$ and $n=0$ are depicted in Figs. \ref{EMwaves} and \ref{Gwaves} correspondingly.
It can be noted that the waveforms of the axial and polar parities remain consistent for varied parameters.
Upon comparing Fig. \ref{EMwaves} and Fig. \ref{Gwaves}, it becomes evident that the gravitational mode demonstrates a slower decay and lower frequency when contrasted with the electromagnetic mode.
As depicted on the right side of Figs. \ref{EMwaves} - \ref{Gwaves}, it can be observed that waveforms with larger values of the parameter $\alpha$ exhibit slower decay and a decrease in frequency.
This suggests that a higher value of $\alpha$ corresponds to smaller imaginary and real parts in the QNM frequencies.
These findings align with the conclusions presented in subsection \ref{QNMs0}.
Additionally, when the cosmological constant $ \Lambda $ is larger, the waveform is more sensitive to variations in the MOG parameter $ \alpha $.
For now, we try to fit $ \psi(t,r) $ using a QNM model comprising a finite number of exponentially damped sinusoids.
Without loss of generality, only considering the fundamental mode for nonrotating black hole, we drop the indices $n$ and $m$.
Hence, we use a modified exponentially decaying function as
\begin{align}
Q(t)=e^{\omega_I t}A_{l}\sin(\omega_R+B_{l}), && t\in (t_0,t_\text{end}),
\end{align}
and choose the range of the fit from $ t_0=120M $ to $ t_\text{end}=160M $.

Last, we present the fitted results in comparison with those calculated by the matrix method in Tabs. $ \ref{tab1} $ and $ \ref{tab2} $, which is electromagnetic and gravitational perturbations, respectively.
Considering the inherent errors that can arise during numerical computations and the limitations imposed by the finite number of parameters in the fitting process, we have confidence in the accuracy of the matrix method based on the obtained fitting results.
\begin{table}[h]
\renewcommand{\arraystretch}{1.2}
\centering
\caption{Contrasting the axial electromagnetic mode for values of $ M = 1 $ and $ \Lambda = 0.03 $.}
\label{tab1}
\setlength\tabcolsep{2mm}{
\begin{tabular}{cccccc}
\hline\hline
\multirow{2}{*}{~\bf }	&
\multirow{1}{*}{~\bf $ \alpha $~}  &
\multicolumn{1}{c}{Matrix method} & \multicolumn{1}{c}{Fitting}
\\
\hline
\multirow{3}{*}{$ l=1 $}
  &$0   $  &0.213721-0.079565i  & 0.213717-0.079568i           \\
  &$0.5 $  &0.128374-0.043767i	& 0.128209-0.043539i           \\
  &$1.0 $  &0.051311-0.016438i	& 0.051816-0.014548i           \\
\\
\multirow{3}{*}{$ l=2 $}
  &$0   $  &0.391893-0.081346i 	&0.391886-0.081368i
\\&$0.5 $  &0.231013-0.044289i	&0.230873-0.044137i
\\&$1.0 $  &0.090893-0.016480i	&0.091239-0.014361i
\\
\hline\hline
\end{tabular}}
\end{table}
\begin{table}[h]
\renewcommand{\arraystretch}{1.2}
\centering
\caption{Contrasting the axial gravitational mode for values of $ M = 1 $ and $ \Lambda = 0.03 $.}
\label{tab2}
\setlength\tabcolsep{2mm}{
\begin{tabular}{cccccc}
\hline\hline
\multirow{2}{*}{~\bf }	&
\multirow{1}{*}{~\bf $ \alpha $~}  &
\multicolumn{1}{c}{Matrix method} & \multicolumn{1}{c}{Fitting}
\\
\hline
\multirow{3}{*}{$ l=2 $}
  &$0   $  &0.319260-0.077710i  &0.319250-0.077717i    \\
  &$0.5 $  &0.174115-0.042734i	&0.174023-0.042581i    \\
  &$1.0 $  &0.065427-0.016336i	&0.066693-0.013963i    \\
\\
\multirow{3}{*}{$ l=3 $}
  &$0   $  &0.512575-0.079989i 	&0.512582-0.080048i
\\&$0.5 $  &0.280741-0.043420i	&0.280594-0.043207i
\\&$1.0 $  &0.106234-0.016389i	&0.105986-0.013908i
\\
\hline\hline
\end{tabular}}
\end{table}

\section{CONCLUSIONS AND EXTENSIONS}\label{Sec.6}
MOG is a covariant modification of GR with a massive vector field $ \phi_{\mu} $ and two scalar fields $ G $ and $ \mu $.
And the vector field is assumed to be of the form $\phi_{\mu}=(\phi_0(r),0,0,0)$, which represents the optimal configuration for a static solution \cite{Moffat:2014aja}.
In this work, we obtained an asymptotic (Anti) de Sitter solution with a cosmological constant in the MOG theory.
Additionally, we have computed the QNM frequencies of electromagnetic and gravitational perturbations for this black hole solution.
Our results show a significant dependence on the dimensionless parameter $\alpha$, and confirm the isospectrality in MOG-dS spacetime if the interaction of matter and the vector field is not considered. We also investigate the consequence of considering the interaction term in calculation of the QNM frequencies. 

In this paper, we only consider the QNMs of MOG-dS spacetime.
However, for the MOG-AdS spacetime, since the AdS/CFT correspondence, the QNM frequencies are still worthy to investigate. 
Another interesting extension is the metric \meq{ds2MOG} can be extended to the rotating situation by assuming a corresponding vector field $\phi_{\mu}=(\phi_0,0,0,\phi_3)$. In rotating case, this vector field will be related to the cosmological constant $ \Lambda $.
Note that in static spacetime, our $\phi_{\mu}$, \eq{phi0}, is equivalent to Eq. (7) in Ref. \cite{Moffat:2014aja} as a special case.
As an extension of this research, our investigation of the MOG-(A)dS spacetime to search for Kerr-like solutions, as well as studying the QNM frequencies and other relevant properties of this black hole, is expected to shed light on the properties of the MOG theory and its applicability in astrophysical scenarios. 
Furthermore, the spins of the merging black holes are expected to play a crucial role in interpreting the aLIGO/Virgo ringdown data accurately. By gaining a better understanding of the spin dynamics in black hole mergers \cite{Jing:2023vzq,Jing:2022vks,Guo:2023niy,Jing:2023vzq}, we can deepen our understanding of the fundamental physics of gravity and the behavior of black holes.

\section{ACKNOWLEDGMENTS}
The author would like to thank Pan Qiyuan, Qin Tan and Wen-Di Guo for their insightful discussions.
This work was partially supported by the National Natural Science Foundation of China under Grants No. 12122504, No. 12375046, No. 12035005, and the Hunan Provincial Natural Science Foundation of China under Grant No. 2022JJ40262.

\appendix

\section{EXPLICIT PERTURBATION EQUATIONS}\label{Appendixeq}
In this appendix, we present all the components of the perturbed Einstein equations:
\begin{equation}\label{EQG1}
\begin{aligned}
&2M^2\alpha(1+\alpha)H_0-2r^3F^3H_2'-r^2F(\lambda-2)K\\
&-F^2\left[2M^2\alpha(1+\alpha)+r^2(\lambda+2F+4rF')\right]\\
&+r^3F(6F+rF')K'+2r^4F^2K''\\
&=4M\sqrt{\alpha}(1+\alpha)\left[F\left(ru_{(1)}'-u_{(1)}\right)+i\omega r u_{(2)}\right]
\end{aligned}
\end{equation}
\begin{equation}\label{EQG2}
\begin{aligned}
-2i\omega rF^2H_2-i\omega r(rF'-2F)K_0\\
+\lambda FH_1+2i\omega r^2 F K'=0,
\end{aligned}
\end{equation}
\begin{equation}\label{EQG3}
\begin{aligned}
i\omega r^3\lambda K+r^3\lambda F'H_1+r^3\lambda F(i\omega H_2+H_1')\\
=4M\sqrt{\alpha}(1+\alpha)\left( \lambda u_{(2)}-Fru_{(3)}' \right),
\end{aligned}
\end{equation}
\begin{equation}\label{EQG4}
\begin{aligned}
\lambda(2F+\lambda-2)h_0-i\omega r \lambda F\left(2h_1+rh_1'\right)\\
-r^2\lambda Fh_0''=-4M\sqrt{\alpha}(1+\alpha)Fu_{(4)}',
\end{aligned}
\end{equation}
\begin{equation}\label{EQG5}
\begin{aligned}
&\left(r^2\lambda+2M^2\alpha(1+\alpha)+2r^3F'\right)H_0\\
&-2r^3FH_0'-4i\omega r^3 FH_1+2r^2F^2\left(r^2\Lambda-1\right)H_2\\
&+r^2\left[2r^2\omega^2-(\lambda-2)F\right]K+r^3F\left(2F+rF'\right)K'\\
&=4M\sqrt{\alpha}(1+\alpha)\left[F\left(ru_{(1)}'-u_{(1)}\right)+i\omega r u_{(2)}\right],
\end{aligned}
\end{equation}
\begin{equation}\label{EQG6}
\begin{aligned}
&r^2\lambda \left(2F+rF'\right)H_0-2r^3\lambda FH_0'-2i\omega r^3 \lambda H_1\\
&-r^2\lambda F^2\left(2F+rF'\right)H_2+2r^3 \lambda F^2 K'\\
&=-8M\sqrt{\alpha}(1+\alpha)F\left(\lambda u_{(1)}+i\omega r u_{(3)}\right)
\end{aligned}
\end{equation}
\begin{equation}\label{EQG7}
\begin{aligned}
&2i\omega r\lambda h_0+\lambda\left[r^2\omega^2-(\lambda-2)F\right]h_1\\
&-i\omega r^2\lambda h_0'=-4i\omega M\sqrt{\alpha}(1+\alpha)u_{(4)},
\end{aligned}
\end{equation}
\begin{equation}\label{EQG8}
\begin{aligned}
i \omega r h_0+rF'Fh_1+rF^2h_1'=0,
\end{aligned}
\end{equation}
\begin{equation}\label{EQG9}
\begin{aligned}
H_0-F^2H_2=0,
\end{aligned}
\end{equation}
\begin{align}\label{EQG10}
&F\left(12M(1+\alpha)(r-M\alpha)+8\Lambda r^4-3\lambda r^2\right)H_0\non
&+3r^4{F'}^2H_0+3r^3F(2F-rF')H_0'+6r^4F^2H_0''\non
&+6i\omega r^3 F(2F+rF')H_1+12i \omega r^4 F^2 H_1'\non
&+F^3\left(12M(1+\alpha)(r-M\alpha)-16\Lambda r^4+3r^2\lambda\right) H_2\non
&+F^2\left(3r^4{F'}^2-6r^4\omega^2\right)H_2+3r^3F^3(2F+rF')H_2'\non
&-6r^4\omega^2FK-6r^3F^2(2F+rF')K'-6r^4F^3K''\non
&=24M\sqrt{\alpha}(1+\alpha)F[F(ru_{(1)}'-u_{(1)})+i\omega ru_{(2)}].
\end{align}

And here are all the components of the perturbed Maxwell equations:
\begin{equation}\label{EQE1}
\begin{aligned}
&i\omega r^2 F' u_{(2)}-i\omega r F[u_{(2)}-u_{(3)}+ru'_{(2)}]\\
&+\lambda Fu_{(1)}+\xi F^2 u_{(1)}-rF^2[\xi u'_{(1)}+ru''_{(1)}]\\
&=\frac{M\sqrt{\alpha}}{2}\Big[(1-2\xi)rF'H_0-2\xi F H_0+F^3rH'_2\\
&~~~-(1-2\xi)F r H'_0-F^2(2rK'-rF'H_2) \Big],
\end{aligned}
\end{equation}
\begin{equation}\label{EQE2}
\begin{aligned}
&(r^2\omega^2/F-\xi F+\xi rF'-\lambda)u_{(2)}+i\omega r u_{(1)}\\
&-i\omega r^2 u'_{(1)}+rF[\xi u'_{(2)}+u'_{(3)}]=M\sqrt{\alpha}\\
&\times \Big[\xi(F-rF')H_1-\xi F rH'_1 \\
&~~~~-\frac{i\omega r}{2}\left(H_0/F+2K_0-FH_2\right)  \Big],
\end{aligned}
\end{equation}
\begin{equation}
\begin{aligned}\label{EQE3}
&F^2r^2u''_{(3)}+F(\xi F+r F')ru'_{(3)}+r^2\omega^2u_{(3)}-i\omega r \lambda u_{(1)}\\
&+\xi F( rF' F-2 F)u_{(3)}+\lambda F[u_{(2)}-ru'_{(2)}]=0,
\end{aligned}
\end{equation}
\begin{equation}\label{EQE4}
\begin{aligned}
&F^2r^2u''_{(4)}+F(\xi F+r F')ru'_{(4)}+r^2\omega^2 u_{(4)}\\
&-F(\lambda+2\xi F-\xi rF')u_{(4)}=\frac{M\sqrt{\alpha}\lambda F}{r}\\
&\times\left[(3\xi-2)h_0-(\xi-1)r h_0'+i\omega r h_1\right].
\end{aligned}
\end{equation}

\section{Explicit coefficients}\label{AppendixCCC}
In this appendix, we present the explicit expressions of the parameters $\varpi$, $c_1$, and $c_2$ as
\begin{equation}
\varpi=3M(1+\alpha)-\frac{2M^2\alpha(1+\alpha)}{r}+\frac{r}{2}\left(\lambda-2\right),
\end{equation}
\begin{equation}
\begin{aligned}
c_1=&\frac{\Lambda}{6}\left[2M(1+\alpha)(4M\alpha-3r)+r^2\lambda\right]\\
&-\frac{M(1+\alpha)}{2r^4}\left[2M^2\alpha(1+\alpha)(4M\alpha-11r)\right.\\
&\left.+Mr^2(12+20\alpha-\alpha \lambda)+\lambda r^3-6r^3\right],\\
\end{aligned}
\end{equation}
\begin{equation}
\begin{aligned}
c_2=&\frac{\Lambda}{3}\left[2M^2\alpha(1+\alpha)+r^2(\lambda-1)\right]-(\lambda-2)\\
&+\frac{M(1+\alpha)}{r^4}\left[2M^2\alpha(1+\alpha)(M\alpha-4r)\right.\\
&\left.+Mr^2(9+12\alpha-\alpha \lambda)+2\lambda r^3-8r^3  \right].
\end{aligned}
\end{equation}

\begin{widetext}
\section{QNM FREQUENCYS TABLES}\label{AppendixQNM}
In this appendix, we will provide some QNMs data for reference.
\begin{table}[h]
\renewcommand{\arraystretch}{1.2}
\centering
\caption{Comparison of the $n=0$, $l=1$ mode electromagnetic QNM frequencies calculated by the matrix method and the WKB approach in the MOG-de Sitter spacetime.}
\label{tab:2}
\setlength\tabcolsep{3mm}{
\begin{tabular}{cccccccc}
\hline\hline
\multirow{2}{*}{~\bf }	&
\multirow{3}{*}{~\bf $ \alpha $~}  &
\multicolumn{2}{c}{Matrix method} & \multicolumn{2}{c}{WKB approach}  \\
\cline{3-4}  \cline{5-6} \cline{7-8}
&~&{axial}&{polar}&{axial}&{polar}\\
\hline
\multirow{4}{*}{$ \Lambda=0.01 $}
&0    &0.237424-0.088396i &0.237424-0.088396i &0.237360-0.088529i &0.237360-0.088529i \\
&0.5  &0.169603-0.057991i &0.169606-0.057990i &0.169566-0.058099i &0.169603-0.058050i \\
&1.0  &0.126914-0.040831i &0.126915-0.040831i &0.126876-0.040914i &0.126921-0.040856i \\
\\
\multirow{4}{*}{$ \Lambda=0.02 $}
  &0    &0.225944-0.084104i &0.225944-0.084104i &0.225891-0.084218i &0.225891-0.084218i
\\&0.5  &0.150578-0.051376i &0.150578-0.051376i &0.150554-0.051446i &0.150573-0.051421i
\\&1.0  &0.097174-0.031146i &0.097174-0.031146i &0.097164-0.031175i &0.097175-0.031161i
\\
\\
\multirow{4}{*}{$ \Lambda=0.03 $}
  &0    &0.213721-0.079565i &0.213721-0.079565i &0.213676-0.079661i &0.213676-0.079661i
\\&0.5  &0.128374-0.043767i &0.128374-0.043767i &0.128361-0.043808i &0.128369-0.043798i
\\&1.0  &0.051311-0.016438i &0.051311-0.016438i &0.051311-0.016444i &0.051311-0.016443i
\\
\hline\hline
\end{tabular}}
\end{table}

\begin{table}[H]
\renewcommand{\arraystretch}{1.2}
\centering
\caption{Comparison of the $n=0$, $l=2$ mode electromagnetic QNM frequencies calculated by the matrix method and the WKB approach in the MOG-de Sitter spacetime.}
\label{tab:3}
\setlength\tabcolsep{3mm}{
\begin{tabular}{cccccccc}
\hline\hline
\multirow{2}{*}{~\bf }	&
\multirow{3}{*}{~\bf $ \alpha $~}  &
\multicolumn{2}{c}{Matrix method} & \multicolumn{2}{c}{WKB approach}  \\
\cline{3-4}  \cline{5-6} \cline{7-8}
&~&{axial}&{polar}&{axial}&{polar}\\
\hline
\multirow{4}{*}{$ \Lambda=0.01 $}
&0    &0.436871-0.090683i &0.436871-0.090683i &0.436859-0.090689i &0.436859-0.090689i \\
&0.5  &0.306868-0.059032i &0.306869-0.059033i &0.306868-0.059037i &0.306868-0.059037i \\
&1.0  &0.226347-0.041330i &0.226347-0.041330i &0.226346-0.041332i &0.226346-0.041332i \\
\\
\multirow{4}{*}{$ \Lambda=0.02 $}
  &0    &0.415023-0.086144i &0.415023-0.086144i &0.415021-0.086149i &0.415021-0.086149i
\\&0.5  &0.271696-0.052168i &0.271696-0.052168i &0.271696-0.052170i &0.271696-0.052170i
\\&1.0  &0.172686-0.031405i &0.172686-0.031405i &0.172686-0.031406i &0.172686-0.031406i
\\
\\
\multirow{4}{*}{$ \Lambda=0.03 $}
  &0    &0.391893-0.081346i &0.391893-0.081346i &0.391892-0.081351i &0.391892-0.081351i
\\&0.5  &0.231013-0.044289i &0.231013-0.044289i &0.231013-0.044290i &0.231013-0.044290i
\\&1.0  &0.090893-0.016480i &0.090893-0.016480i &0.090893-0.016480i &0.090893-0.016480i
\\
\hline\hline
\end{tabular}}
\end{table}

\begin{table}[H]
\renewcommand{\arraystretch}{1.2}
\centering
\caption{Comparison of the $n=1$, $l=2$ mode electromagnetic QNM frequencies calculated by the matrix method and the WKB approach in the MOG-de Sitter spacetime.}
\label{tab:4}
\setlength\tabcolsep{3mm}{
\begin{tabular}{cccccccc}
\hline\hline
\multirow{2}{*}{~\bf }	&
\multirow{3}{*}{~\bf $ \alpha $~}  &
\multicolumn{2}{c}{Matrix method} & \multicolumn{2}{c}{WKB approach}  \\
\cline{3-4}  \cline{5-6} \cline{7-8}
&~&{axial}&{polar}&{axial}&{polar}\\
\hline
\multirow{4}{*}{$ \Lambda=0.01 $}
&0    &0.418304-0.276619i &0.418304-0.276619i &0.418336-0.276648i &0.418336-0.276648i \\
&0.5  &0.297394-0.179206i &0.297398-0.179205i &0.297400-0.179221i &0.297401-0.179220i \\
&1.0  &0.221090-0.124987i &0.221089-0.124989i &0.221088-0.124997i &0.221088-0.124996i \\
\\
\multirow{4}{*}{$ \Lambda=0.02 $}
  &0    &0.399004-0.262002i &0.399004-0.262002i &0.398998-0.262020i &0.398998-0.262020i
\\&0.5  &0.264994-0.157671i &0.264995-0.157671i &0.264995-0.157679i &0.264995-0.157679i
\\&1.0  &0.170291-0.094475i &0.170291-0.094475i &0.170291-0.094479i &0.170291-0.094479i
\\
\\
\multirow{4}{*}{$ \Lambda=0.03 $}
  &0    &0.378322-0.246730i &0.378322-0.246730i &0.378315-0.246737i &0.378315-0.246737i
\\&0.5  &0.226858-0.133378i &0.226858-0.133378i &0.226859-0.133384i &0.226859-0.133384i
\\&1.0  &0.090546-0.049449i &0.090546-0.049449i &0.090546-0.049449i &0.090546-0.049449i
\\
\hline\hline
\end{tabular}}
\end{table}

\begin{table}[H]
\renewcommand{\arraystretch}{1.2}
\centering
\caption{Comparison of the $n=0$, $l=2$ mode gravitational QNM frequencies calculated by the matrix method and the WKB approach in the MOG-de Sitter spacetime.}
\label{tab:5}
\setlength\tabcolsep{3mm}{
\begin{tabular}{cccccccc}
\hline\hline
\multirow{2}{*}{~\bf }	&
\multirow{3}{*}{~\bf $ \alpha $~}  &
\multicolumn{2}{c}{Matrix method} & \multicolumn{2}{c}{WKB approach}  \\
\cline{3-4}  \cline{5-6} \cline{7-8}
&~&{axial}&{polar}&{axial}&{polar}\\
\hline
\multirow{4}{*}{$ \Lambda=0.01 $}
&0    &0.356480-0.085468i &0.356476-0.085488i &0.356446-0.085425i &0.356508-0.085460i \\
&0.5  &0.232415-0.055130i &0.232416-0.055131i &0.232406-0.055089i &0.232420-0.055135i \\
&1.0  &0.164552-0.038807i &0.164552-0.038807i &0.164557-0.038779i &0.164551-0.038817i \\
\\
\multirow{4}{*}{$ \Lambda=0.02 $}
  &0    &0.338390-0.081755i &0.338391-0.081756i &0.338373-0.081694i &0.338414-0.081730i
\\&0.5  &0.205305-0.049545i &0.205305-0.049545i &0.205306-0.049513i &0.205308-0.049547i
\\&1.0  &0.124959-0.030350i &0.124959-0.030350i &0.124964-0.030339i &0.124959-0.030353i	
\\
\\
\multirow{4}{*}{$ \Lambda=0.03 $}
  &0    &0.319260-0.077710i &0.319261-0.077710i &0.319253-0.077654i &0.319278-0.077689i
\\&0.5  &0.174115-0.042734i &0.174115-0.042734i &0.174120-0.042715i &0.174117-0.042735i
\\&1.0  &0.065427-0.016336i &0.065427-0.016336i &0.065427-0.016335i &0.065427-0.016336i	
\\
\hline\hline
\end{tabular}}
\end{table}

\begin{table}[H]
\renewcommand{\arraystretch}{1.2}
\centering
\caption{Comparison of the $n=1$, $l=2$ mode gravitational QNM frequencies calculated by the matrix method and the WKB approach in the MOG-de Sitter spacetime.}
\label{tab:6}
\setlength\tabcolsep{3mm}{
\begin{tabular}{cccccccc}
\hline\hline
\multirow{2}{*}{~\bf }	&
\multirow{3}{*}{~\bf $ \alpha $~}  &
\multicolumn{2}{c}{Matrix method} & \multicolumn{2}{c}{WKB approach}  \\
\cline{3-4}  \cline{5-6} \cline{7-8}
&~&{axial}&{polar}&{axial}&{polar}\\
\hline
\multirow{4}{*}{$ \Lambda=0.01 $}
&0    &0.333886-0.261280i &0.333178-0.261547i &0.332943-0.261414i &0.333259-0.261778i \\
&0.5  &0.219868-0.167799i &0.219824-0.167828i &0.219685-0.167606i &0.219813-0.167962i \\
&1.0  &0.157582-0.117468i &0.157584-0.117470i &0.157552-0.117308i &0.157580-0.117575i \\
\\
\multirow{4}{*}{$ \Lambda=0.02 $}
  &0    &0.318786-0.249097i &0.318747-0.249178i &0.318534-0.248812i &0.318762-0.249137i
\\&0.5  &0.197007-0.149752i &0.197007-0.149753i &0.196953-0.149573i &0.197009-0.149810i
\\&1.0  &0.122207-0.091234i &0.122207-0.091234i &0.122209-0.091169i &0.122209-0.091260i
\\
\\
\multirow{4}{*}{$ \Lambda=0.03 $}
  &0    &0.303029-0.235892i &0.303032-0.235886i &0.302884-0.235548i &0.303040-0.235830i
\\&0.5  &0.169344-0.128600i &0.169344-0.128600i &0.169331-0.128489i &0.169346-0.128621i
\\&1.0  &0.065074-0.049012i &0.065074-0.049012i &0.065076-0.049009i &0.064886-0.049157i
\\
\hline\hline
\end{tabular}}
\end{table}

\begin{table}[H]
\renewcommand{\arraystretch}{1.2}
\centering
\caption{Comparison of the $n=0$, $l=3$ mode gravitational QNM frequencies calculated by the matrix method and the WKB approach in the MOG-de Sitter spacetime.}
\label{tab:7}
\setlength\tabcolsep{3mm}{
\begin{tabular}{cccccccc}
\hline\hline
\multirow{2}{*}{~\bf }	&
\multirow{3}{*}{~\bf $ \alpha $~}  &
\multicolumn{2}{c}{Matrix method} & \multicolumn{2}{c}{WKB approach}  \\
\cline{3-4}  \cline{5-6} \cline{7-8}
&~&{axial}&{polar}&{axial}&{polar}\\
\hline
\multirow{4}{*}{$ \Lambda=0.01 $}
&0    &0.571997-0.088726i &0.571994-0.088723i &0.571991-0.088720i &0.571991-0.088720i \\
&0.5  &0.373753-0.056874i &0.373753-0.056874i &0.373753-0.056874i &0.373753-0.056874i \\
&1.0  &0.265549-0.039795i &0.265549-0.039795i &0.265549-0.039795i &0.265549-0.039795i \\
\\
\multirow{4}{*}{$ \Lambda=0.02 $}
  &0    &0.543115-0.084496i &0.543115-0.084496i &0.543115-0.084495i &0.543115-0.084496i
\\&0.5  &0.330558-0.050710i &0.330558-0.050710i &0.330558-0.050711i &0.330558-0.050711i
\\&1.0  &0.202230-0.030753i &0.202230-0.030753i &0.202230-0.030754i &0.202230-0.030754i
\\
\\
\multirow{4}{*}{$ \Lambda=0.03 $}
  &0    &0.512575-0.079989i &0.512575-0.079989i &0.512576-0.079988i &0.512575-0.079989i
\\&0.5  &0.280741-0.043420i &0.280741-0.043420i &0.280741-0.043420i &0.280741-0.043420i
\\&1.0  &0.106234-0.016389i &0.106234-0.016389i &0.106234-0.016389i &0.106232-0.016389i
\\
\hline\hline
\end{tabular}}
\end{table}

\begin{table}[H]
\renewcommand{\arraystretch}{1.2}
\centering
\caption{Comparison of the $n=1$, $l=3$ mode gravitational QNM frequencies calculated by the matrix method and the WKB approach in the MOG-de Sitter spacetime.}
\label{tab:8}
\setlength\tabcolsep{3mm}{
\begin{tabular}{cccccccc}
\hline\hline
\multirow{2}{*}{~\bf }	&
\multirow{3}{*}{~\bf $ \alpha $~}  &
\multicolumn{2}{c}{Matrix method} & \multicolumn{2}{c}{WKB approach}  \\
\cline{3-4}  \cline{5-6} \cline{7-8}
&~&{axial}&{polar}&{axial}&{polar}\\
\hline
\multirow{4}{*}{$ \Lambda=0.01 $}
&0    &0.557466-0.268766i &0.557476-0.268715i &0.557454-0.268637i &0.557454-0.268639i \\
&0.5  &0.365916-0.171794i &0.365915-0.171791i &0.365914-0.171790i &0.365914-0.171792i \\
&1.0  &0.261177-0.119917i &0.261177-0.119917i &0.261177-0.119917i &0.261177-0.119918i \\
\\
\multirow{4}{*}{$ \Lambda=0.02 $}
  &0    &0.530746-0.255365i &0.530745-0.255364i &0.530743-0.255358i &0.530743-0.255359i
\\&0.5  &0.325270-0.152718i &0.325270-0.152718i &0.325271-0.152718i &0.325271-0.152719i
\\&1.0  &0.200410-0.092375i &0.200410-0.092375i &0.200410-0.092375i &0.200410-0.092376i
\\
\\
\multirow{4}{*}{$ \Lambda=0.03 $}
  &0    &0.502255-0.241334i &0.502255-0.241334i &0.502254-0.241330i &0.502254-0.241331i
\\&0.5  &0.277603-0.130496i &0.277603-0.130496i &0.277603-0.130496i &0.277603-0.130496i
\\&1.0  &0.105988-0.049171i &0.105988-0.049171i &0.105988-0.049171i &0.105965-0.049182i
\\
\hline\hline
\end{tabular}}
\end{table}
\end{widetext}


\begin{thebibliography}{000}

\bibitem{LIGOScientific:2016sjg}
B.~P.~Abbott \textit{et al.} [LIGO Scientific and Virgo],
``GW151226: Observation of Gravitational Waves from a 22-Solar-Mass Binary Black Hole Coalescence,''
Phys. Rev. Lett. \textbf{116}, no.24, 241103 (2016)
doi:10.1103/PhysRevLett.116.241103
[arXiv:1606.04855 [gr-qc]].

\bibitem{Bozzola:2020mjx}
G.~Bozzola and V.~Paschalidis,
``General Relativistic Simulations of the Quasicircular Inspiral and Merger of Charged Black Holes: GW150914 and Fundamental Physics Implications,''
Phys. Rev. Lett. \textbf{126} (2021) no.4, 041103
doi:10.1103/PhysRevLett.126.041103
[arXiv:2006.15764 [gr-qc]].

\bibitem{Gupta:2021rod}
P. K. Gupta, T. F. M. Spieksma, P. T. H. Pang, G. Koekoek and C. V. Broeck,
``Bounding dark charges on binary black holes using gravitational waves,''
Phys. Rev. D \textbf{104} (2021) no.6, 063041
doi:10.1103/PhysRevD.104.063041
[arXiv:2107.12111 [gr-qc]].

\bibitem{Carullo:2021oxn}
G.~Carullo, D.~Laghi, N.~K.~Johnson-McDaniel, W.~Del Pozzo, O.~J.~C.~Dias, M.~Godazgar and J.~E.~Santos,
``Constraints on Kerr-Newman black holes from merger-ringdown gravitational-wave observations,''
Phys. Rev. D \textbf{105} (2022) no.6, 062009
doi:10.1103/PhysRevD.105.062009
[arXiv:2109.13961 [gr-qc]].


\bibitem{Gibbons:1975kk}
G.~W.~Gibbons,
``Vacuum Polarization and the Spontaneous Loss of Charge by Black Holes,''
Commun. Math. Phys. \textbf{44} (1975), 245-264
doi:10.1007/BF01609829

\bibitem{Blandford:1977ds}
R.~D.~Blandford and R.~L.~Znajek,
``Electromagnetic extractions of energy from Kerr black holes,''
Mon. Not. Roy. Astron. Soc. \textbf{179} (1977), 433-456
doi:10.1093/mnras/179.3.433

\bibitem{Cardoso:2016olt}
V.~Cardoso, C.~F.~B.~Macedo, P.~Pani and V.~Ferrari,
``Black holes and gravitational waves in models of minicharged dark matter,''
JCAP \textbf{05} (2016), 054
[erratum: JCAP \textbf{04} (2020), E01]
doi:10.1088/1475-7516/2016/05/054
[arXiv:1604.07845 [hep-ph]].


\bibitem{Moffat:2005si}
J.~W.~Moffat,
``Scalar-tensor-vector gravity theory,''
JCAP \textbf{03}, 004 (2006)
doi:10.1088/1475-7516/2006/03/004
[arXiv:gr-qc/0506021 [gr-qc]].

\bibitem{Moffat:2013sja}
J.~W.~Moffat and S.~Rahvar,
``The MOG weak field approximation and observational test of galaxy rotation curves,''
Mon. Not. Roy. Astron. Soc. \textbf{436}, 1439-1451 (2013)
doi:10.1093/mnras/stt1670
[arXiv:1306.6383 [astro-ph.GA]].

\bibitem{Moffat:2013uaa}
J.~W.~Moffat and S.~Rahvar,
``The MOG weak field approximation \textendash{} II. Observational test of $Chandra$ X-ray clusters,''
Mon. Not. Roy. Astron. Soc. \textbf{441}, no.4, 3724-3732 (2014)
doi:10.1093/mnras/stu855
[arXiv:1309.5077 [astro-ph.CO]].



\bibitem{Moffat:2014bfa}
J.~W.~Moffat,
``Structure Growth and the CMB in Modified Gravity (MOG),''
[arXiv:1409.0853 [astro-ph.CO]].

\bibitem{Moffat:2014pia}
J.~W.~Moffat and V.~T.~Toth,
``Rotational velocity curves in the Milky Way as a test of modified gravity,''
Phys. Rev. D \textbf{91}, no.4, 043004 (2015)
doi:10.1103/PhysRevD.91.043004
[arXiv:1411.6701 [astro-ph.GA]].


\bibitem{Moffat:2014aja}
J.~W.~Moffat,
``Black Holes in Modified Gravity (MOG),''
Eur. Phys. J. C \textbf{75}, no.4, 175 (2015)
doi:10.1140/epjc/s10052-015-3405-x
[arXiv:1412.5424 [gr-qc]].

\bibitem{Lee:2017fbq}
H.~C.~Lee and Y.~J.~Han,
``Innermost stable circular orbit of Kerr-MOG black hole,''
Eur. Phys. J. C \textbf{77}, no.10, 655 (2017)
doi:10.1140/epjc/s10052-017-5152-7
[arXiv:1704.02740 [gr-qc]].

\bibitem{Qiao:2020fta}
X.~Qiao, M.~Wang, Q.~Pan and J.~Jing,
``Kerr-MOG black holes with stationary scalar clouds,''
Eur. Phys. J. C \textbf{80}, no.6, 509 (2020)
doi:10.1140/epjc/s10052-020-8062-z


\bibitem{Moffat:2015kva}
J.~W.~Moffat,
``Modified Gravity Black Holes and their Observable Shadows,''
Eur. Phys. J. C \textbf{75}, no.3, 130 (2015)
doi:10.1140/epjc/s10052-015-3352-6
[arXiv:1502.01677 [gr-qc]].

\bibitem{Guo:2018kis}
M.~Guo, N.~A.~Obers and H.~Yan,
``Observational signatures of near-extremal Kerr-like black holes in a modified gravity theory at the Event Horizon Telescope,''
Phys. Rev. D \textbf{98} (2018) no.8, 084063
doi:10.1103/PhysRevD.98.084063
[arXiv:1806.05249 [gr-qc]].

\bibitem{Wang:2018prk}
H.~M.~Wang, Y.~M.~Xu and S.~W.~Wei,
``Shadows of Kerr-like black holes in a modified gravity theory,''
JCAP \textbf{03}, 046 (2019)
doi:10.1088/1475-7516/2019/03/046
[arXiv:1810.12767 [gr-qc]].

\bibitem{Qin2022}
X.~Qin, S.~Chen, Z.~Zhang and J.~Jing,
``Polarized Image of a Rotating Black Hole in Scalar\textendash{}Tensor\textendash{}Vector\textendash{}Gravity Theory,''
Astrophys. J. \textbf{938}, no.1, 2 (2022)
doi:10.3847/1538-4357/ac8f49
[arXiv:2207.12034 [gr-qc]].

\bibitem{Rahvar:2022yhj}
S.~Rahvar,
``Hamiltonian formalism for dynamics of particles in MOG,''
Mon. Not. Roy. Astron. Soc. \textbf{514} (2022) no.3, 4601-4605
doi:10.1093/mnras/stac1560
[arXiv:2206.02453 [gr-qc]].

\bibitem{Rouhani:2023qzy}
S.~Rouhani and S.~Rahvar,
``MOG as symmetry breaking in Scalar-Vector-Tensor gravity,''
[arXiv:2308.13511 [gr-qc]].


\bibitem{Planck:2018nkj}
N.~Aghanim \textit{et al.} [Planck],
``Planck 2018 results. I. Overview and the cosmological legacy of Planck,''
Astron. Astrophys. \textbf{641} (2020), A1
doi:10.1051/0004-6361/201833880
[arXiv:1807.06205 [astro-ph.CO]].

\bibitem{Maldacena:1997re}
J.~M.~Maldacena,
``The Large N limit of superconformal field theories and supergravity,''
Adv. Theor. Math. Phys. \textbf{2} (1998), 231-252
doi:10.4310/ATMP.1998.v2.n2.a1
[arXiv:hep-th/9711200 [hep-th]].

\bibitem{Nunez:2003eq}
A.~Nunez and A.~O.~Starinets,
``AdS / CFT correspondence, quasinormal modes, and thermal correlators in N=4 SYM,''
Phys. Rev. D \textbf{67} (2003), 124013
doi:10.1103/PhysRevD.67.124013
[arXiv:hep-th/0302026 [hep-th]].

\bibitem{Son:2007vk}
D.~T.~Son and A.~O.~Starinets,
``Viscosity, Black Holes, and Quantum Field Theory,''
Ann. Rev. Nucl. Part. Sci. \textbf{57} (2007), 95-118
doi:10.1146/annurev.nucl.57.090506.123120
[arXiv:0704.0240 [hep-th]].

\bibitem{Hartnoll:2009sz}
S.~A.~Hartnoll,
``Lectures on holographic methods for condensed matter physics,''
Class. Quant. Grav. \textbf{26} (2009), 224002
doi:10.1088/0264-9381/26/22/224002
[arXiv:0903.3246 [hep-th]].

\bibitem{Herzog:2009xv}
C.~P.~Herzog,
``Lectures on Holographic Superfluidity and Superconductivity,''
J. Phys. A \textbf{42} (2009), 343001
doi:10.1088/1751-8113/42/34/343001
[arXiv:0904.1975 [hep-th]].






\bibitem{Regge:1957td}
T.~Regge and J.~A.~Wheeler,
``Stability of a Schwarzschild singularity,''
Phys. Rev. \textbf{108}, 1063-1069 (1957)
doi:10.1103/PhysRev.108.1063

\bibitem{Zerilli:1970se}
F.~J.~Zerilli,
``Effective potential for even parity Regge-Wheeler gravitational perturbation equations,''
Phys. Rev. Lett. \textbf{24}, 737-738 (1970)
doi:10.1103/PhysRevLett.24.737

\bibitem{Zerilli:1970wzz}
F.~J.~Zerilli,
``Gravitational field of a particle falling in a schwarzschild geometry analyzed in tensor harmonics,''
Phys. Rev. D \textbf{2}, 2141-2160 (1970)
doi:10.1103/PhysRevD.2.2141


\bibitem{Liu2023}
W.~Liu, X.~Fang, J.~Jing and A.~Wang,
``Gauge invariant perturbations of general spherically symmetric spacetimes,''
Sci. China Phys. Mech. Astron. \textbf{66}, no.1, 210411 (2023)
doi:10.1007/s11433-022-1956-4
[arXiv:2201.01259 [gr-qc]].


\bibitem{Chandrasekhar:1984siy}
S.~Chandrasekhar,
``The Mathematical Theory of Black Holes,''
Fundam. Theor. Phys. \textbf{9}, 5-26 (1984)
doi:10.1007/978-94-009-6469-3\_2

\bibitem{Berti:2009kk}
E.~Berti, V.~Cardoso and A.~O.~Starinets,
``Quasinormal modes of black holes and black branes,''
Class. Quant. Grav. \textbf{26}, 163001 (2009)
doi:10.1088/0264-9381/26/16/163001
[arXiv:0905.2975 [gr-qc]].



\bibitem{Sheoran:2017dwb}
P.~Sheoran, A.~Herrera-Aguilar and U.~Nucamendi,
``Mass and spin of a Kerr black hole in modified gravity and a test of the Kerr black hole hypothesis,''
Phys. Rev. D \textbf{97}, no.12, 124049 (2018)
doi:10.1103/PhysRevD.97.124049
[arXiv:1712.03344 [gr-qc]].


\bibitem{Zhao:2023jiz}
Y.~Zhao, W.~Liu, C.~Zhang, X.~Fang and J.~Jing,
``The Quasinormal Modes and Isospectrality of Bardeen (Anti-) de Sitter Black Holes,''
[arXiv:2306.02332 [gr-qc]].

\bibitem{Rosa:2011my}
J.~G.~Rosa and S.~R.~Dolan,
``Massive vector fields on the Schwarzschild spacetime: quasi-normal modes and bound states,''
Phys. Rev. D \textbf{85}, 044043 (2012)
doi:10.1103/PhysRevD.85.044043
[arXiv:1110.4494 [hep-th]].

\bibitem{Zhang:2023wwk}
X.~Zhang, M.~Wang and J.~Jing,
``Quasinormal modes and late time tails of perturbation fields on a Schwarzschild-like black hole with a global monopole in the Einstein-bumblebee theory,''
Sci. China Phys. Mech. Astron. \textbf{66} (2023) no.10, 100411
doi:10.1007/s11433-023-2153-6
[arXiv:2307.10856 [gr-qc]].

\bibitem{Pani:2013ija}
P.~Pani, E.~Berti and L.~Gualtieri,
``Gravitoelectromagnetic Perturbations of Kerr-Newman Black Holes: Stability and Isospectrality in the Slow-Rotation Limit,''
Phys. Rev. Lett. \textbf{110} (2013) no.24, 241103
doi:10.1103/PhysRevLett.110.241103
[arXiv:1304.1160 [gr-qc]].

\bibitem{Pani:2013wsa}
P.~Pani, E.~Berti and L.~Gualtieri,
``Scalar, Electromagnetic and Gravitational Perturbations of Kerr-Newman Black Holes in the Slow-Rotation Limit,''
Phys. Rev. D \textbf{88} (2013), 064048
doi:10.1103/PhysRevD.88.064048
[arXiv:1307.7315 [gr-qc]].

\bibitem{Nomura:2020tpc}
K.~Nomura, D.~Yoshida and J.~Soda,
``Stability of magnetic black holes in general nonlinear electrodynamics,''
Phys. Rev. D \textbf{101} (2020) no.12, 124026
doi:10.1103/PhysRevD.101.124026
[arXiv:2004.07560 [gr-qc]].

\bibitem{Meng:2022oxg}
K.~Meng and S.~J.~Zhang,
``Gravito-Electromagnetic Perturbations and QNMs of Regular Black Holes,''
[arXiv:2210.00295 [gr-qc]].

\bibitem{Guo:2022rms}
W.~D.~Guo, Q.~Tan and Y.~X.~Liu,
``Gravito-Electromagnetic coupled perturbations and quasinormal modes of a charged black hole with scalar hair,''
[arXiv:2212.08784 [gr-qc]].


\bibitem{Thorne:1980ru}
K.~S.~Thorne,
``Multipole Expansions of Gravitational Radiation,''
Rev. Mod. Phys. \textbf{52} (1980), 299-339
doi:10.1103/RevModPhys.52.299

\bibitem{Zerilli:1974ai}
F.~J.~Zerilli,
``Perturbation analysis for gravitational and electromagnetic radiation in a reissner-nordstroem geometry,''
Phys. Rev. D \textbf{9}, 860-868 (1974)
doi:10.1103/PhysRevD.9.860

\bibitem{Moncrief:1974am}
V.~Moncrief,
``Gravitational perturbations of spherically symmetric systems. I. The exterior problem.,''
Annals Phys. \textbf{88}, 323-342 (1974)
doi:10.1016/0003-4916(74)90173-0

\bibitem{Moncrief:1975sb}
V.~Moncrief,
``Gauge-invariant perturbations of Reissner-Nordstrom black holes,''
Phys. Rev. D \textbf{12} (1975), 1526-1537
doi:10.1103/PhysRevD.12.1526

\bibitem{Cardoso:2001bb}
V.~Cardoso and J.~P.~S.~Lemos,
``Quasinormal modes of Schwarzschild anti-de Sitter black holes: Electromagnetic and gravitational perturbations,''
Phys. Rev. D \textbf{64} (2001), 084017
doi:10.1103/PhysRevD.64.084017
[arXiv:gr-qc/0105103 [gr-qc]].

\bibitem{Zhidenko:2003wq}
A.~Zhidenko,
``Quasinormal modes of Schwarzschild de Sitter black holes,''
Class. Quant. Grav. \textbf{21}, 273-280 (2004)
doi:10.1088/0264-9381/21/1/019
[arXiv:gr-qc/0307012 [gr-qc]].

\bibitem{Lin:2016sch}
K.~Lin and W.~L.~Qian,
``A Matrix Method for Quasinormal Modes: Schwarzschild Black Holes in Asymptotically Flat and (Anti-) de Sitter Spacetimes,''
Class. Quant. Grav. \textbf{34}, no.9, 095004 (2017)
doi:10.1088/1361-6382/aa6643
[arXiv:1610.08135 [gr-qc]].

\bibitem{Lin:2017oag}
K.~Lin, W.~L.~Qian, A.~B.~Pavan and E.~Abdalla,
``A matrix method for quasinormal modes: Kerr and Kerr\textendash{}Sen black holes,''
Mod. Phys. Lett. A \textbf{32}, no.25, 1750134 (2017)
doi:10.1142/S0217732317501346
[arXiv:1703.06439 [gr-qc]].

\bibitem{Lin:2019mmf}
K.~Lin and W.~L.~Qian,
``On matrix method for black hole quasinormal modes,''
Chin. Phys. C \textbf{43}, no.3, 035105 (2019)
doi:10.1088/1674-1137/43/3/035105
[arXiv:1902.08352 [gr-qc]].

\bibitem{Lei:2021kqv}
Y.~Lei, M.~Wang and J.~Jing,
``Maxwell perturbations in a cavity with Robin boundary conditions: two branches of modes with spectrum bifurcation on Schwarzschild black holes,''
Eur. Phys. J. C \textbf{81}, no.12, 1129 (2021)
doi:10.1140/epjc/s10052-021-09942-8
[arXiv:2108.04146 [gr-qc]].

\bibitem{Liu:2022dcn}
W.~Liu, X.~Fang, J.~Jing and J.~Wang,
``QNMs of slowly rotating Einstein-Bumblebee Black Hole,''
Eur. Phys. J. C \textbf{83}, 83 (2023)
doi:10.1140/epjc/s10052-023-11231-5
[arXiv:2211.03156 [gr-qc]].

\bibitem{Schutz:1985km}
B.~F.~Schutz and C.~M.~Will,
``BLACK HOLE NORMAL MODES: A SEMIANALYTIC APPROACH,''
Astrophys. J. Lett. \textbf{291}, L33-L36 (1985)
doi:10.1086/184453

\bibitem{Iyer:1986np}
S.~Iyer and C.~M.~Will,
``Black Hole Normal Modes: A {WKB} Approach. 1. Foundations and Application of a Higher Order {WKB} Analysis of Potential Barrier Scattering,''
Phys. Rev. D \textbf{35}, 3621 (1987)
doi:10.1103/PhysRevD.35.3621

\bibitem{Konoplya:2003ii}
R.~A.~Konoplya,
``Quasinormal behavior of the d-dimensional Schwarzschild black hole and higher order WKB approach,''
Phys. Rev. D \textbf{68}, 024018 (2003)
doi:10.1103/PhysRevD.68.024018
[arXiv:gr-qc/0303052 [gr-qc]].

\bibitem{Berti:2005ys}
E.~Berti, V.~Cardoso and C.~M.~Will,
``On gravitational-wave spectroscopy of massive black holes with the space interferometer LISA,''
Phys. Rev. D \textbf{73} (2006), 064030
doi:10.1103/PhysRevD.73.064030
[arXiv:gr-qc/0512160 [gr-qc]].


\bibitem{Ciftci:2005xn}
H.~Ciftci, R.~L.~Hall and N.~Saad,
``Perturbation theory in a framework of iteration methods,''
Phys. Lett. A \textbf{340}, 388-396 (2005)
doi:10.1016/j.physleta.2005.04.030
[arXiv:math-ph/0504056 [math-ph]].


\bibitem{Manfredi:2017xcv}
L.~Manfredi, J.~Mureika and J.~Moffat,
``Quasinormal Modes of Modified Gravity (MOG) Black Holes,''
Phys. Lett. B \textbf{779}, 492-497 (2018)
doi:10.1016/j.physletb.2017.11.006
[arXiv:1711.03199 [gr-qc]].

\bibitem{Brito:2018hjh}
R.~Brito and C.~Pacilio,
``Quasinormal modes of weakly charged Einstein-Maxwell-dilaton black holes,''
Phys. Rev. D \textbf{98}, no.10, 104042 (2018)
doi:10.1103/PhysRevD.98.104042
[arXiv:1807.09081 [gr-qc]].

\bibitem{Wei:2018aft}
S.~W.~Wei and Y.~X.~Liu,
``Merger estimates for rotating Kerr black holes in modified gravity,''
Phys. Rev. D \textbf{98}, no.2, 024042 (2018)
doi:10.1103/PhysRevD.98.024042
[arXiv:1803.09530 [gr-qc]].

\bibitem{isoLoop}
D.~del-Corral and J.~Olmedo,
``Breaking of isospectrality of quasinormal modes in nonrotating loop quantum gravity black holes,''
Phys. Rev. D \textbf{105}, no.6, 064053 (2022)
doi:10.1103/PhysRevD.105.064053
[arXiv:2201.09584 [gr-qc]].

\bibitem{isoChSi}
S.~Bhattacharyya and S.~Shankaranarayanan,
``Distinguishing general relativity from Chern-Simons gravity using gravitational wave polarizations,''
Phys. Rev. D \textbf{100}, no.2, 024022 (2019)
doi:10.1103/PhysRevD.100.024022
[arXiv:1812.00187 [gr-qc]].

\bibitem{isoLovelock}
C.~B.~Prasobh and V.~C.~Kuriakose,
``Quasinormal Modes of Lovelock Black Holes,''
Eur. Phys. J. C \textbf{74}, no.11, 3136 (2014)
doi:10.1140/epjc/s10052-014-3136-4
[arXiv:1405.5334 [gr-qc]].

\bibitem{Leaver1985}
E.~W.~Leaver,
``An Analytic representation for the quasi normal modes of Kerr black holes,''
Proc. Roy. Soc. Lond. A \textbf{402} (1985), 285-298
doi:10.1098/rspa.1985.0119
\bibitem{Pani:2013pma}
P.~Pani,
``Advanced Methods in Black-Hole Perturbation Theory,''
Int. J. Mod. Phys. A \textbf{28} (2013), 1340018
doi:10.1142/S0217751X13400186
[arXiv:1305.6759 [gr-qc]].

\bibitem{Gundlach:1993tp}
C.~Gundlach, R.~H.~Price and J.~Pullin,
``Late time behavior of stellar collapse and explosions: 1. Linearized perturbations,''
Phys. Rev. D \textbf{49} (1994), 883-889
doi:10.1103/PhysRevD.49.883
[arXiv:gr-qc/9307009 [gr-qc]].

\bibitem{Abdalla:2010nq}
E.~Abdalla, C.~E.~Pellicer, J.~de Oliveira and A.~B.~Pavan,
``Phase transitions and regions of stability in Reissner-Nordstr\"om holographic superconductors,''
Phys. Rev. D \textbf{82} (2010), 124033
doi:10.1103/PhysRevD.82.124033
[arXiv:1010.2806 [hep-th]].

\bibitem{Zhu:2014sya}
Z.~Zhu, S.~J.~Zhang, C.~E.~Pellicer, B.~Wang and E.~Abdalla,
``Stability of Reissner-Nordstr\"om black hole in de Sitter background under charged scalar perturbation,''
Phys. Rev. D \textbf{90} (2014) no.4, 044042
doi:10.1103/PhysRevD.90.044042
[arXiv:1405.4931 [hep-th]].

\bibitem{Lin:2022owb}
K.~Lin and W.~L.~Qian,
``Echoes in star quasinormal modes using an alternative finite difference method,''
[arXiv:2204.09531 [gr-qc]].

\bibitem{Fu:2022cul}
G.~Fu, D.~Zhang, P.~Liu, X.~M.~Kuang, Q.~Pan and J.~P.~Wu,
``Quasinormal modes and Hawking radiation of a charged Weyl black hole,''
Phys. Rev. D \textbf{107} (2023) no.4, 044049
doi:10.1103/PhysRevD.107.044049
[arXiv:2207.12927 [gr-qc]].

\bibitem{Tan:2022vfe}
Q.~Tan, W.~D.~Guo and Y.~X.~Liu,
``Sound from extra dimensions: Quasinormal modes of a thick brane,''
Phys. Rev. D \textbf{106} (2022) no.4, 044038
doi:10.1103/PhysRevD.106.044038
[arXiv:2205.05255 [gr-qc]].


\bibitem{Jing:2022vks}
J.~Jing, S.~Long, W.~Deng, M.~Wang and J.~Wang,
``New self-consistent effective one-body theory for spinless binaries based on the post-Minkowskian approximation,''
Sci. China Phys. Mech. Astron. \textbf{65} (2022) no.10, 100411
doi:10.1007/s11433-022-1951-1
[arXiv:2208.02420 [gr-qc]].


\bibitem{Jing:2023okh}
J.~Jing, W.~Deng, S.~Long and J.~Wang,
``Effective metric of spinless binaries with radiation-reaction effect up to fourth post-Minkowskian order in effective-one-body theory,''
Eur. Phys. J. C \textbf{83} (2023) no.7, 608
[erratum: Eur. Phys. J. C \textbf{83} (2023) no.8, 712]
doi:10.1140/epjc/s10052-023-11705-6
[arXiv:2307.05971 [gr-qc]].

\bibitem{Jing:2023vzq}
J.~Jing, W.~Deng, S.~Long and J.~Wang,
``Self-consistent effective-one-body theory for spinning binaries based on post-Minkowskian approximation,''
Sci. China Phys. Mech. Astron. \textbf{66} (2023) no.7, 270411
doi:10.1007/s11433-023-2084-1
[arXiv:2305.03225 [gr-qc]].

\bibitem{Guo:2023niy}
Y.~Guo, H.~Nakajima and W.~Lin,
``Gravitational-wave equation in effective one-body background for spinless binary,''
Sci. China Phys. Mech. Astron. \textbf{66} (2023) no.7, 270412
doi:10.1007/s11433-023-2087-8
[arXiv:2301.08318 [gr-qc]].
































\end{thebibliography}
\end{document}